\documentclass[useAMS,usenatbib]{mn2e}

\usepackage[latin1]{inputenc}
\usepackage{graphicx}
\usepackage{epstopdf}

\usepackage{subfigure}
\usepackage{rotate}
\usepackage{amsmath, float}
\usepackage{txfonts}
\usepackage{fleqn}
\usepackage{color}
\usepackage{comment}

\usepackage{mathrsfs}  
\usepackage{booktabs}
\usepackage{colortbl}

\newcolumntype{G}{>{\columncolor[gray]{0.8}}l} 

\newcommand{\be}{\begin{equation}}
\newcommand{\ee}{\end{equation}}
\newcommand{\bdm}{\begin{displaymath}}
\newcommand{\edm}{\end{displaymath}}
\newcommand{\bea}{\begin{multline}}
\newcommand{\eea}{\end{multline}}
\newcommand{\ba}{\begin{align}}
\newcommand{\ea}{\end{align}}

\renewcommand{\sigma}{\varpi^2}

\def\simlt{\mathrel{\hbox{\rlap{\hbox{\lower4pt\hbox{$\sim$}}}\hbox{$<$}}}}
\def\simgt{\mathrel{\hbox{\rlap{\hbox{\lower4pt\hbox{$\sim$}}}\hbox{$>$}}}}

\title[Investigating GR equilibria of magnetized NSs]
{The role of currents distribution in general relativistic equilibria of magnetized neutron stars}
\author[N. Bucciantini, A.~G. Pili, L. Del Zanna]{
N. Bucciantini$^{1,2}$\thanks{E-mail: niccolo@arcetri.astro.it}
A.~G. Pili$^{3,1,2}$, L. Del Zanna$^{3,1,2}$ \\
$^{1}$INAF - Osservatorio Astrofisico di Arcetri, Largo E. Fermi 5, I-50125 Firenze, Italy\\
$^{2}$INFN - Sezione di Firenze, Via G. Sansone 1, I-50019 Sesto F.~no  (Firenze), Italy\\
$^{3}$Dipartimento di Fisica e Astronomia, Universit\`a degli Studi di
Firenze, Via G. Sansone 1, I-50019 Sesto F.~no  (Firenze), Italy}

\begin{document}
 
\date{Accepted / Received}

\maketitle

\label{firstpage}

\begin{abstract}
Magnetic fields play a critical role in the phenomenology of neutron
stars. There is virtually no observable aspect which is not governed
by them. Despite this, only recently efforts have been done to model
magnetic fields in the correct general relativistic regime,
characteristic of these compact objects. In this work we present, for
the first time a comprehensive and detailed 
parameter study, in general relativity, of the role that
the current distribution, and the related magnetic field structure,
have in determining the precise structure of neutron stars.
In particular, we show how the
presence of localized currents can modify the field strength at the
stellar surface, and we look for general trends, both in terms of
energetic properties, and magnetic field configurations. 
Here we verify that, among other things,  
for a large class of different current distributions the resulting magnetic
configurations are always dominated by the poloidal component of the current.

\end{abstract}


\section{Introduction}

The peculiar phenomenology of Anomalous X-Ray Pulsars (AXPs) and Soft Gamma
Repeaters (SGRs) \citep{Mereghetti08a}, has led to the introduction of a new class of
astrophysical sources, called \textit{magnetars} \citep{Duncan_Thompson92,Thompson_Duncan93}. 
These are Neutron Stars (NSs) endowed with a very strong magnetic field $\sim
10^{14} - 10^{15}$~G, about two orders of magnitude stronger than
what is found in normal pulsars. This strong magnetic field is supposed
to form at birth, involving possibly some form of dynamo action \citep{Bonanno_Rezzolla+03a,Rheinhardt_Geppert05a}. At
birth the star is still fluid, and will remain so for a typical
Kelvin-Helmoholtz timescale [$\sim 10-100$ s \citep{Pons_Reddy+99a}], before the formation of
a crust. This timescale is much longer than a typical Alfv\'en
timescale, and one expects the magnetic field, in the end, to relax to some
form of stable or meta-stable configuration 
\citep{Braithwaite_Spruit06a,Braithwaite_Norlund06a,Braithwaite09a}.

It is well known that it is the magnetic field that dictates how NSs manifest themselves
in the electromagnetic spectrum. In the astrophysical community, a vast effort has been devoted
to model the outer magnetosphere of these objects, from the
pioneering work of \citet{Goldreich_Julian68} to the most recent numerical models by
\citet{Tchekhovskoy_Spitkovsky13a}. 
This contrasts with the attention that has been placed on modeling
the interior structure of these objects, which has been mostly driven
by key fundamental questions of nuclear and theoretical physics. Most
of the efforts in this respect have gone toward the study of their
Equation of State (EoS) \citep{Chamel_Haensel08a,Lattimer12a} and their cooling
properties \citep{Yakovlev_Pethick04a,Yakovlev_Gnedin+05a}.

When dealing with problems related to the EoS or the cooling, one can
safely assume that the NS is spherically symmetric (which is a good
approximation except for the fastest rotators). This greatly
simplifies the equations that one needs to solve. 
On the other hand, it is obvious that the geometry of the magnetic field is, instead,
a truly multidimensional problem. This means that, except for trivial cases, the
task  of finding equilibrium solutions can only be handled numerically. 
The techniques to model magnetic field in General Relativity,
i.e. GR-MHD and its extensions \citep{Font08a}, have been mostly
developed in the last 10-15 years. As a consequence, only
recently attention has been paid to the study of magnetic field
structure and evolution in NSs \citep{Bocquet_Bonazzola+95a,Haskell_Samuellsson+08a,Konno01a,Kiuchi_Yoshida08a,Kiuchi_Kotake+09a,Frieben_Rezzolla12a,Yazadjiev12a,Fujisawa_Yoshida+12a,Yoshida_Yoshida+06a,Ciolfi_ferrari+09a,Ciolfi_Ferrari+10a,Ciolfi_Rezzolla13a,Lander_Jones09a,Glampedakis_Andersson+12a,Pili_Bucciantini+14a,Pili_Bucciantini+14b,Armaza_Reissenegger+14a}. 

In general, these studies have focused on trying to obtain specific
configurations, and on investigating a few key aspects of the field
morphology. However, a detailed study of the parameter space is still
partially lacking. This, obviously, raises questions about the robustness and
generality of some conclusions. Moreover, it does not allow us to
understand if there are general trends and expectations. 

For this reason, in this paper we present a detailed parameter study in GR-MHD
of currents distribution and the related magnetic field configurations in NSs. 
Extending, in part, the work done in \citet{Pili_Bucciantini+14a},
we introduce several new functional forms for the current distribution,
and investigate the properties of the resulting magnetic field. We
show that there are general trends associated with the nonlinear
behavior  of currents, both in terms of global integrated quantities
and for what concerns the structure of the magnetic field at the
surface. We also show the interplay of various nonlinear terms, and
address the robustness of several outcomes.

This paper is organized as follows. In Sect.~\ref{sec:math} we
introduce the formalism, and in particular in Sect.~\ref{sec:freefun} the
various functional forms used for the currents are discussed. 
In Sect.~\ref{sec:results} we present our results, 
and finally conclude in Sect.~\ref{sec:conclusion}.

\section{The mathematical framework}
\label{sec:math}

Let us briefly describe here our mathematical framework, in
particular the equations that are solved to derive the magnetic field
equilibrium configurations, and the choices we have adopted in
order to describe different magnetic geometries.
For further details the reader is referred to \citet{Pili_Bucciantini+14a},
where the full mathematical setup is presented.
In the present work we will assume that the magnetic field strength is well
below the energy equipartition value, so that the field does not
affect neither the overall fluid configuration (assumed to be static and barotropic)
nor the metric, as discussed in  \citet{Pili_Bucciantini+14b}.
Thus, given a consistent spacetime metric and fluid configuration,
we are going to see how to build a magnetic field configuration on top of it.

In the following we assume a signature $(-,+,+,+)$ for the spacetime
metric. Quantities are expressed in geometrized units $c= G =1$, unless otherwise
stated, and all $\sqrt{4\upi}$ factors are absorbed in the definition of the electromagnetic fields.

\subsection{The Grad-Shafranov equation}
\label{sec:gs}

The spacetime and matter distribution of a nonrotating NS, under the
assumption of a negligible magnetic field, are spherically symmetric. In
this case  the spacetime is also \textit{conformally flat}, and the line 
element can be  written, using standard
isotropic coordinates $(t,r,\theta,\phi)$, as
\be
{\rm d}s^2 = - \alpha^2 {\rm d}t^2 + \psi^4 ({\rm d}r^2+ r^2 {\rm d}\theta^2+r^2 \sin^2 \! \theta\,
{\rm d}\phi^2),
\label{eq:iso}
\ee
where $\alpha$ is the \textit{lapse function} and $\psi$ is the \textit{conformal
factor}, which are functions only of the radial coordinate $r$.
These metric terms are related to the matter distribution by two
elliptic equations, derived from Einstein equations, namely
\be
\Delta \psi = [ - 2\upi \psi^5 e] ,
\label{eq:xcfc_psi}
\ee
\be
\Delta (\alpha\psi) = [ 2\upi (e + 6p)\psi^{4} ]  (\alpha\psi),
\label{eq:xcfc_alpha}
\ee
where $\Delta$ is the usual 3D Laplacian of flat spacetime, while
$e$ and $p$ represent, respectively,
the energy density and the pressure  measured by
the Eulerian observer. For a more general discussion in the case of
rotation and/or a strong magnetic field see
\citep{Bucciantini_Del-Zanna11a,Pili_Bucciantini+14a,Pili_Bucciantini+14b}.

For an axisymmetric configuration  in a static spacetime, the electromagnetic field can be described 
uniquely in terms of a magnetic potential that coincides with the $\phi$-component
of the vector potential $A_\phi$ , which is usually referred to as the \textit{magnetic flux
function}. In particular the solenoidality condition, together with  axisymmetry,
allows one to express the poloidal component of the magnetic field as the gradient of a magnetic flux
function, whereas any toroidal component at equilibrium must be related to $A_\phi$ by means of a 
scalar \textit{current function} $\mathcal{I}$, that depends on
$A_\phi$ alone. Thus, the components of the magnetic field are given by
\be
B^r = \frac{ \upartial_\theta A_\phi}{\psi^6 r^2\sin\theta}, \quad
B^\theta = -  \frac{ \upartial_r A_\phi}{\psi^6 r^2\sin\theta},\quad
B^\phi = \frac{\mathcal{I}(A_\phi)}{\alpha\psi^4 r^2\sin^2\theta},
\label{eq:aphi}
\ee
for any choice of the \emph{free} function $\mathcal{I}(A_\phi)$.
Notice that here we are implicitly assuming the presence of a nonvanishing
poloidal component. In the case of purely toroidal magnetic
field, on the other hand, the vector potential $A_\phi$ is not defined and a different
approach must be adopted [see \citet{Pili_Bucciantini+14a} for details].

The Euler equation for a static and barotropic (the pressure is a
function of rest mass density alone) GRMHD configuration
can be written as 
\be
\upartial_i \ln h + \upartial_i \ln\alpha  = \frac{{\rm d}\mathcal{M}}{{\rm d} A_\phi}\upartial_i A_\phi,
\label{eq:euler}
\ee
where $i=r,\theta$, leading to the magnetic \textit{Bernoulli-like integral}
\be
\ln{\left( \frac{h}{h_{\rm c}}\right)} + \ln{\left(
    \frac{\alpha}{\alpha_{\rm c}}\right)} - \mathcal{M} = 0,
\label{eq:bernoulli}
\ee
in which the label ${\rm c}$ refers to values at the centre of the NS.
Here $\rho$ is the rest mass density, $h:=(e+p)/\rho$ is the specific enthalpy,
and the Lorentz force has been written in terms of the
gradient of the \textit{magnetization function} $\mathcal{M}$, our second
scalar function of $A_\phi$.
In other words, if $J^i= \alpha^{-1} \epsilon^{ijk} \upartial_j(\alpha B_k)$ is the 
conduction current, then
\be
\rho h \upartial_i \mathcal{M} = \epsilon_{ijk}J^j B^k.
\label{eq:lorentzf}
\ee
This implies that the current components can be expressed
in terms of the functions $\mathcal{I}(A_\phi)$ and
$\mathcal{M}(A_\phi)$ as
\be
J^r = \alpha^{-1} B^r \frac{{\rm d}\mathcal{I}}{{\rm d} A_\phi}, \quad
J^\theta = \alpha^{-1} B^\theta \frac{{\rm d}\mathcal{I}}{{\rm d} A_\phi},\quad
J^\phi = \rho h \, \frac{{\rm d}\mathcal{M}}{{\rm d} A_\phi} + 
\frac{\mathcal{I}}{\sigma}\frac{{\rm d}\mathcal{I}}{{\rm d} A_\phi},
\label{eq:cur}
\ee
where we have introduced the generalized cylindrical radius $\varpi := \alpha \psi^2 r\sin\!\theta$.

The magnetic flux function $A_\phi$ is related to the metric terms and the hydrodynamical
quantities through the so-called \emph{Grad-Shafranov} (GS) equation \citep{Pili_Bucciantini+14a}:
\be
\tilde{\Delta}_3 \tilde{A}_\phi
+  \frac{\upartial A_\phi \upartial\ln (\alpha\psi^{-2})}{r \sin\theta}
+ \psi^8 r \sin\!\theta \left( \rho h \frac{{\rm d} \mathcal{M}}{{\rm d} A_\phi}
+ \frac{\mathcal{I}}{\sigma}\frac{{\rm d}\mathcal{I}}{{\rm d}A_\phi} \right) = 0
\label{eq:gs}
\ee
which is obtained by working out the derivatives of the magnetic field in
Eq.~\eqref{eq:lorentzf}. For convenience we introduced
the regularized potential $\tilde{A}_\phi=A_\phi / (r\sin\theta)$,
and the following differential operators
\be
\tilde{\Delta}_3  \! := \!  \Delta - \frac{1}{r^2\sin^2\!\theta}  \! = \! 
\upartial^2_r + \frac{2}{r}\upartial_r+\frac{1}{r^2}\upartial_\theta^2
+ \frac{1}{r^2\tan{\theta}}\upartial_\theta - \frac{1}{r^2\sin^2\!\theta},
\ee
\be
\upartial f\upartial g := \upartial_r f \upartial_r g+\frac{1}{r^2}\upartial_\theta f \upartial_\theta g .
\ee
The GS equation, which governs the hydromagnetic equilibrium inside the star, can be
also extended outside it, just by neglecting the term proportional to
the rest mass density.
In this case one recovers the force-free regime \citep{Pili_Bucciantini+14c}.

\subsection{The currents distribution}
\label{sec:freefun}

As anticipated, the purpose of this paper is to study the properties of various
distributions of induction currents, and related magnetic field
configurations, in order to derive general trends and to
understand what parameter governs the key aspects of the magnetic field
geometry. For this reason we have adopted functional forms for the free
$\mathcal{M}(A_\phi)$ and $\mathcal{I}(A_\phi)$ functions allowing us to sample a large
parameter space, and to investigate configurations with important
morphological properties.

The magnetization function $\mathcal{M}$ has been chosen in order to
include nonlinear terms in the form
\be
\mathcal{M}(A_\phi)=k_{\rm pol} A_\phi \left[1 +\frac{\xi}{\nu+1}
\left(\frac{A_\phi}{A_{\phi}^{\rm max}}\right)^{\nu}\right],
\label{eq:mbern}
\ee
where  $k_{\rm pol}$ is the \emph{poloidal magnetization constant},
 $\nu$ is the \emph{poloidal magnetization index} of the nonlinear
 term, which is normalized to the maximum value of the vector
 potential itself $A_{\phi}^{\rm max}$. This normalization prevents the
 nonlinear term from diverging, and it allows the system to remain within
 the weak field limit. The magnetization function $\mathcal{M}$
 vanishes outside the surface of the NS.

For the current function $\mathcal{I}$, we adopted either the form
\be
\mathcal{I}(A_\phi)=\frac{a}{\zeta+1}\Theta [A_\phi-A_\phi^{\rm sur}] 
\frac{(A_\phi-A_\phi^{\rm sur})^{\zeta+1}}{{(A_\phi^{\rm sur})}^{\zeta}},
\label{eq:fbern1}
\ee
or
\be
\mathcal{I}(A_\phi)=\frac{a}{\zeta+1}\Theta [A_\phi-A_\phi^{\rm sur}] 
\frac{(A_\phi-A_\phi^{\rm sur})^{\zeta+1}(A_\phi^{\rm max}-A_\phi)^{\zeta+1}}{{(A_\phi^{\rm sur}A_\phi^{\rm max})}^{\zeta+1/2}},
\label{eq:fbern2}
\ee
where $\Theta [.]$ is the Heaviside function, $A_\phi^{\rm sur}$ is
the maximum value that the $\phi$ component of the vector potential
reaches on the stellar surface, $a$ is the \emph{toroidal magnetization constant} and
$\zeta$ is the \emph{toroidal magnetization index}. In both cases the
toroidal magnetic field is fully confined within the star. However,
the first case corresponds to a
\emph{Twisted Torus} (TT) configuration, where the azimuthal current has the same
sign over its domain and the toroidal field reaches its maximum where the poloidal
field vanishes. 
On the other hand, the second case corresponds to a \emph{Twisted Ring} (TR) configuration, 
where the current changes its sign, and the toroidal field vanishes in the same
place where the poloidal field goes to zero. 

With our normalization choices in Eqs.~\eqref{eq:mbern}
\eqref{eq:fbern1} \eqref{eq:fbern2}, the solution of the GS
equation does not depend on the strength of the magnetic
field. In the limit of a weak field, the metric and fluid quantities
$\rho$, $h$, $\psi$, and $\alpha$ can be assumed as fixed, such that, 
if  $A_\phi$ is a solution of the GS equation
for  given values of $k_{\rm pol}, \xi, \nu, a, \zeta$,
then $\eta A_\phi$ will be a solution of the GS equation for $\eta k_{\rm
  pol}, \xi, \nu, a, \zeta$, for both TT and TR cases. 
Our solution can be thus renormalized to
any value of the magnetic field strength. In particular we have chosen
 to display our solution by normalizing the strength of the magnetic
field at the pole to $10^{14}$~G. We verified that the limit
of a weak field holds to a high level of accuracy up to a maximum strength 
$\sim 10^{16}$~G, corresponding to a typical surface magnetic field $\sim$ a
few $\times 10^{15}$~G (Pili et al. 2014 submitted). For higher
fields, we observe non-linear variations in the ratios of magnetic quantities, higher
than the overall accuracy of our scheme (see
Sect.~\ref{sec:numerical}). A partial investigation of the strong field
regime, for purely poloidal cases, is presented in Appendix~\ref{sec:strong}.


\subsection{The numerical setup}
\label{sec:numerical}

In all our models we assume that the NS is described by a polytropic
EoS $p = K_{\rm a}\rho^{\gamma_{\rm a}}$ (a special case of the
general barotropic EoS $p=p(\rho)$), with an adiabatic index
$\gamma_{\rm a} = 2$. Unless otherwise stated, the polytropic constant
is chosen to be $K_{\rm a} = 110$ (in geometrized units\footnote{This
  corresponds to $K_{\rm a} = 1.6 \times 10^5$
  cm$^5$g$^{-1}$s$^{-2}$}). Given that we are here interested in
studying different distributions of currents, in the limit of weak magnetic
fields, we assume a fiducial model for the NS that is unaffected by the
magnetic field itself, unless otherwise stated. The role of a strong
field has been already investigated in a previous paper
\citep{Pili_Bucciantini+14a}.

The fiducial NS model used for computing, on top of it, the
various magnetic field configurations, has  a central rest mass density 
$\rho_{\rm c}=6.354 \times 10 ^{14} \, \mbox{g} \, \mbox{cm}^{-3}$,  a
baryonic mass $M_0=1.500 M_{\sun}$, a gravitational mass $M=1.400 M_{\sun}$, 
and a circumferential radius $R_{\rm circ}=15.22 \, \mbox{km}$
(corresponding to an isotropic radius $R_{\rm NS} = 13.06 \, \mbox{km}$). 
This fiducial model is computed in isotropic coordinates using the
algorithms and the numerical scheme described in
\citet{Bucciantini_Del-Zanna11a}.

On top of this fiducial model we then solve the GS equation, 
Eq.~\eqref{eq:gs}, as described in \citet{Pili_Bucciantini+14a}. Let us
briefly recall here the main features of the algorithm employed.
The GS equation is a nonlinear vector Poisson
equation for the azimuthal component of the regularized potential
$\tilde{A}_{\phi}$.  The solution is searched in the form of a series of vector spherical
harmonics
\be
\tilde{A}_\phi
(r,\theta):=\sum_{l=0}^{l_{\rm max}}[C_l(r)Y^\prime_l(\theta)],
\label{eq:harmonics}
\ee 
where $Y_l(\theta)$ are the standard scalar spherical
harmonics, the $^\prime$ indicates derivation with respect to $\theta$, and
we have used the axisymmetric assumption to exclude terms with $m \neq 0$. 
We want to stress here again that in the low magnetization limit one can safely assume that 
the metric and the matter distribution are unaffected by the magnetic field, then
in the present approximation we only need to solve the GS equation, for
any given fluid structure and associated spacetime metric.

Our algorithm allows us to solve  the GS equation over the entire
numerical domain including both the interior of the star and  the surrounding magnetosphere 
where the rest mass density is numerically set to a fiducial small value ($\sim 10^{-6}$ times the value of the
rest mass density at the centre of the star), without the need of a matching
procedure between the exterior solution with the interior one. The
threshold value for the rest mass density is chosen such that lowering it
further produces negligible changes (much smaller than the overall accuracy
of our scheme). We want also to point here that typical rest mass densities in
the atmospheres of proto-NS are $\sim 10^6 -10^8$ g cm$^{-3}$ \citep{Thompson_Burrows+01a}. This
also guarantees  smoothness of the solutions at the stellar surface, avoiding surface currents.
Moreover, the harmonic decomposition ensures the correct behavior of the solution
on the axis of symmetry,  and the asymptotic trend of the radial
coefficients $C_l(r)$ can be correctly imposed such that they go to
0 with parity $(-1)^l$ at the centre, and as $C_l(r)\propto
r^{-(l+1)}$ at the outer boundary.

The decomposition in Eq.~\eqref{eq:harmonics} reduces the GS
equation to a system of radial 2nd-order elliptical nonlinear
PDEs for the various coefficients $C_l(r)$. These are solved,
 using a direct tridiagonal matrix inversion
\citep{Bucciantini_Del-Zanna11a,Pili_Bucciantini+14a}. The entire
procedure is repeated until the solution converges with accuracy $\sim 10^{-8}$.

The numerical solutions presented here are always computed in a spherical
domain  covering the range $r=[0,22]$ and $\theta=[0,\upi]$. A uniform
spherical grid is adopted with 700 points in the radial direction and
400 in the angular one. The harmonic decomposition of the vector
potential has $l_{\rm max}=60$.  We verified that the overall accuracy of the solutions
is $\sim 10^{-3}$.

\section{Numerical equilibrium models}
\label{sec:results}

\subsection{Purely poloidal configurations}
\label{sec:poloidal}

Let us begin by illustrating the properties of models with a purely
poloidal field, obtained by a purely toroidal current with $\mathcal{I}=0$,
so that all properties will be determined by the function $\mathcal{M}(A_\phi)$ alone. 
As already pointed out in our previous paper
\citep{Pili_Bucciantini+14a}, the parity of the magnetic field with respect to the
equator depends on the parity of the linear current term in the
magnetization function $\mathcal{M}$. This is
proportional to the rest mass density $\rho$, it is symmetric with respect to
the equator, whereas the related magnetic field is antisymmetric. The
nonlinear term cannot change this parity. The result is that all our
models are \textit{dipole-dominated}, and only terms odd in $l$ are
present in Eq.~\eqref{eq:harmonics}. See Appendix~\ref{sec:anti} for a
discussion on how one can obtain antisymmetric solutions.

The value of $\xi$ can be chosen such that the nonlinear term in
Eq.~\eqref{eq:mbern}, leads
to subtractive currents ($\xi < 0$) or additive currents ($\xi >
0$), while the value of $\nu$ sets how much concentrated this current is.

\begin{figure*}
	\centering
	\includegraphics[width=.33\textwidth, bb=0 20 452 440, clip]{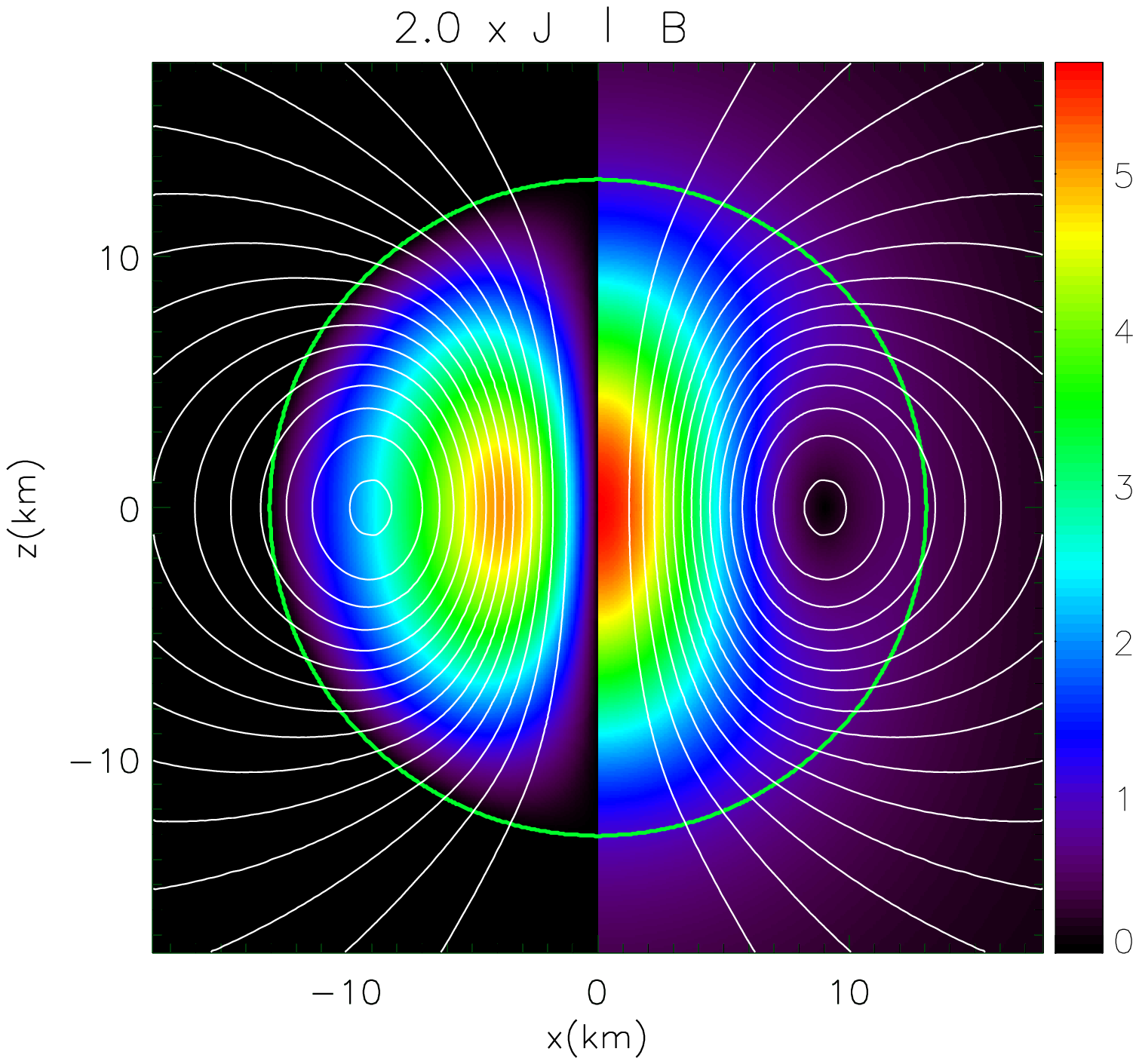}
	\includegraphics[width=.33\textwidth, bb=0 20 452 440, clip]{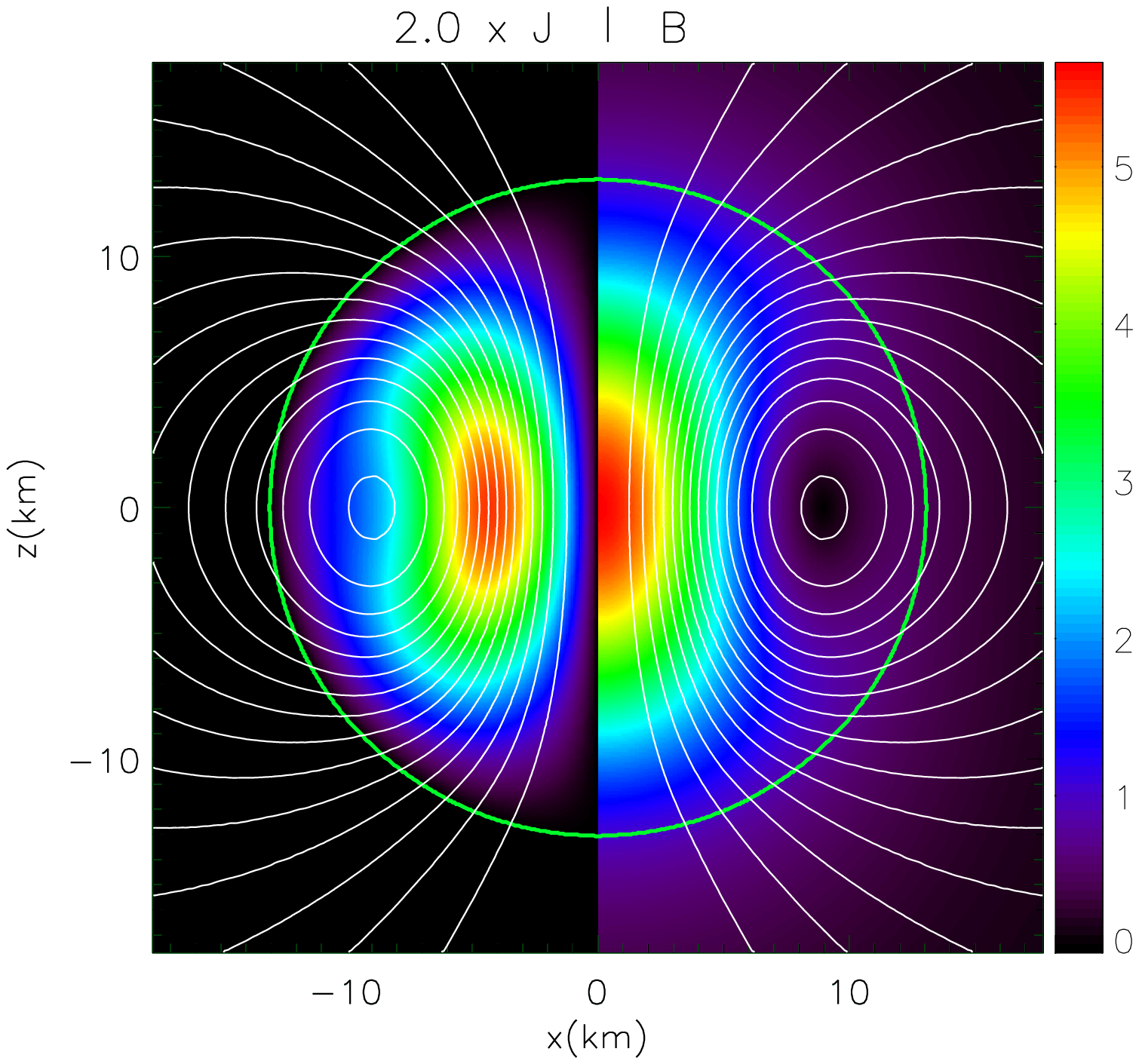}
	\includegraphics[width=.33\textwidth, bb=0 20 452 440, clip]{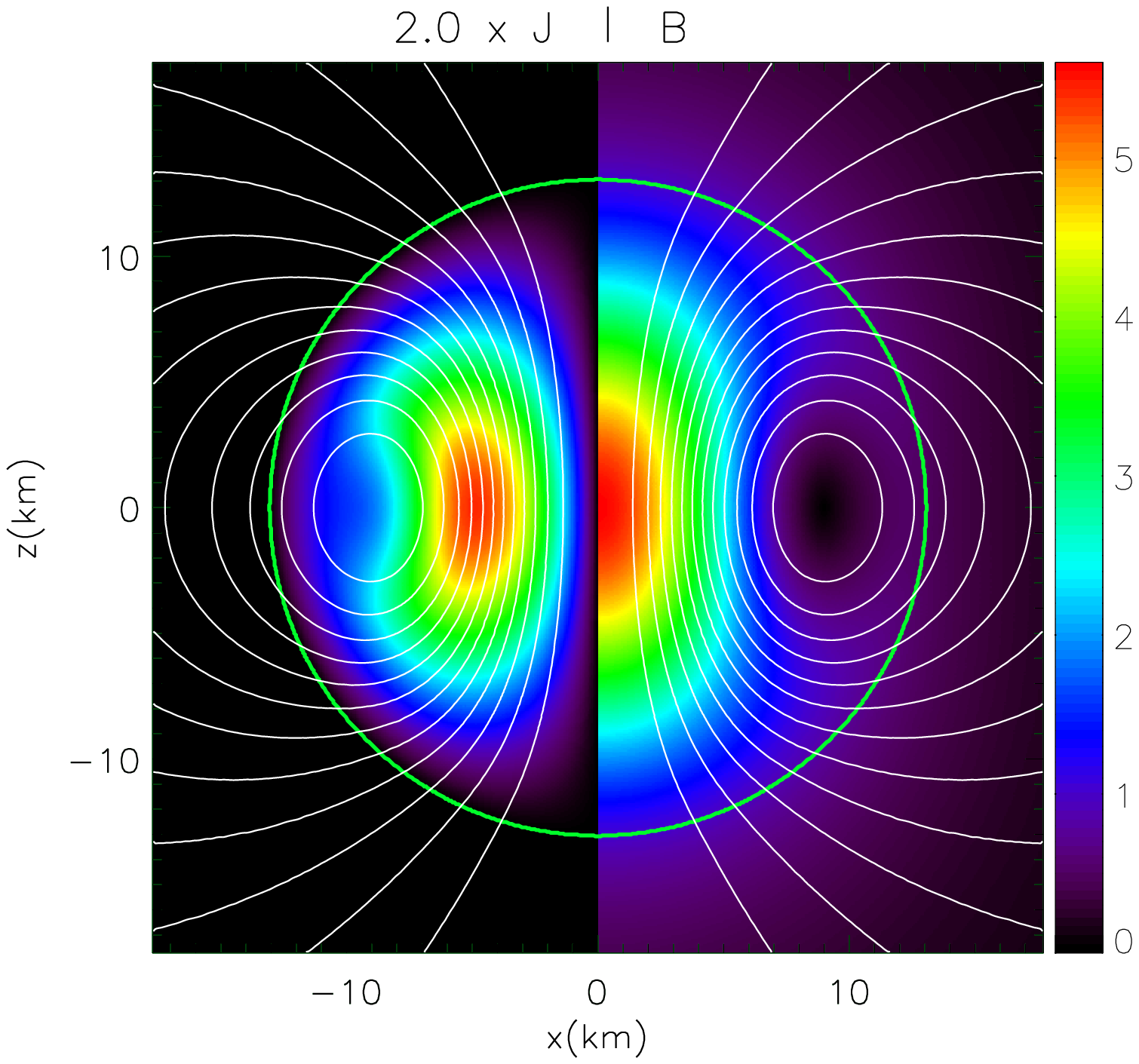}\\
    \includegraphics[width=.33\textwidth, bb=0 20 452 440, clip]{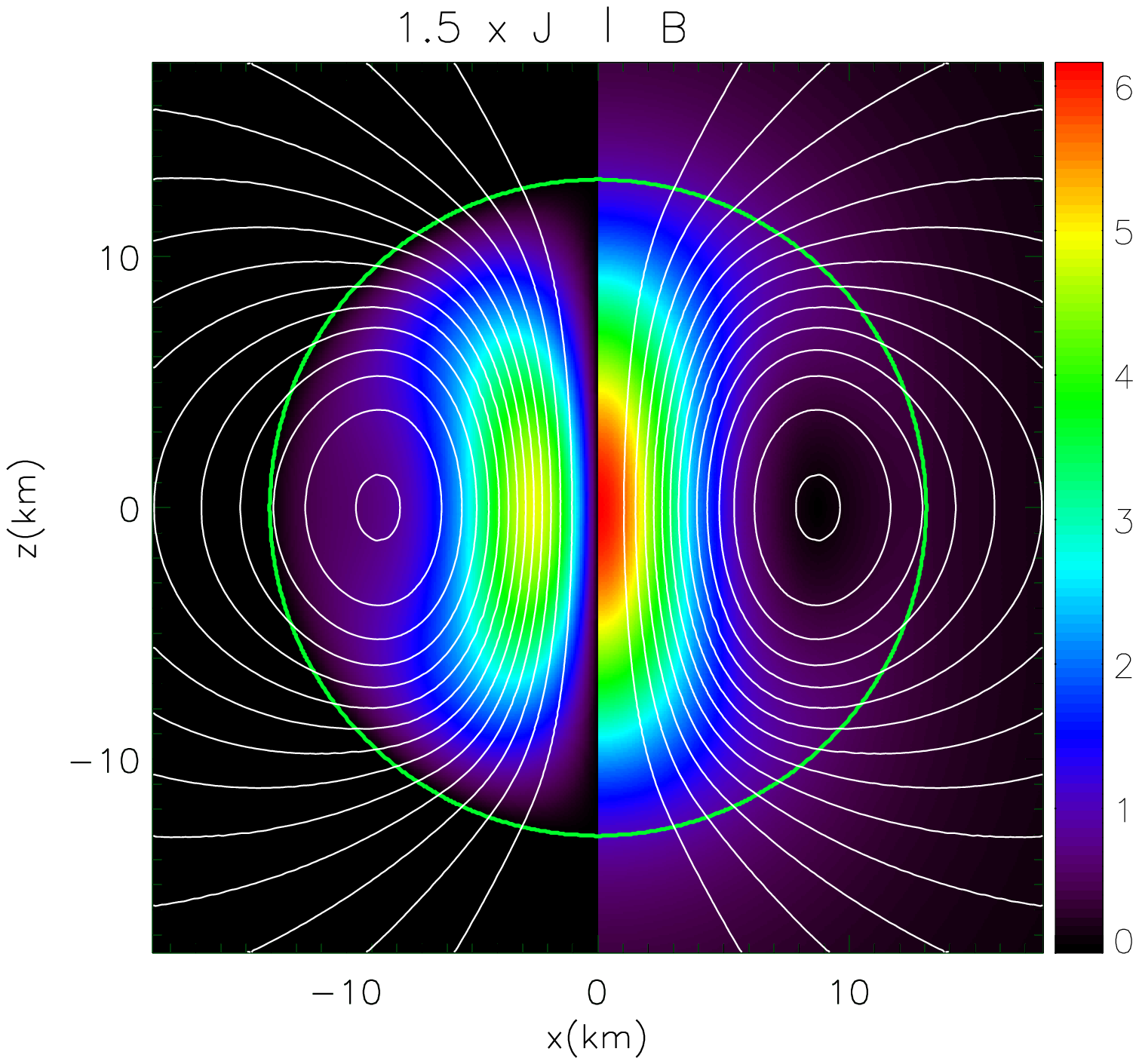}
	\includegraphics[width=.33\textwidth, bb=0 20 452 440, clip]{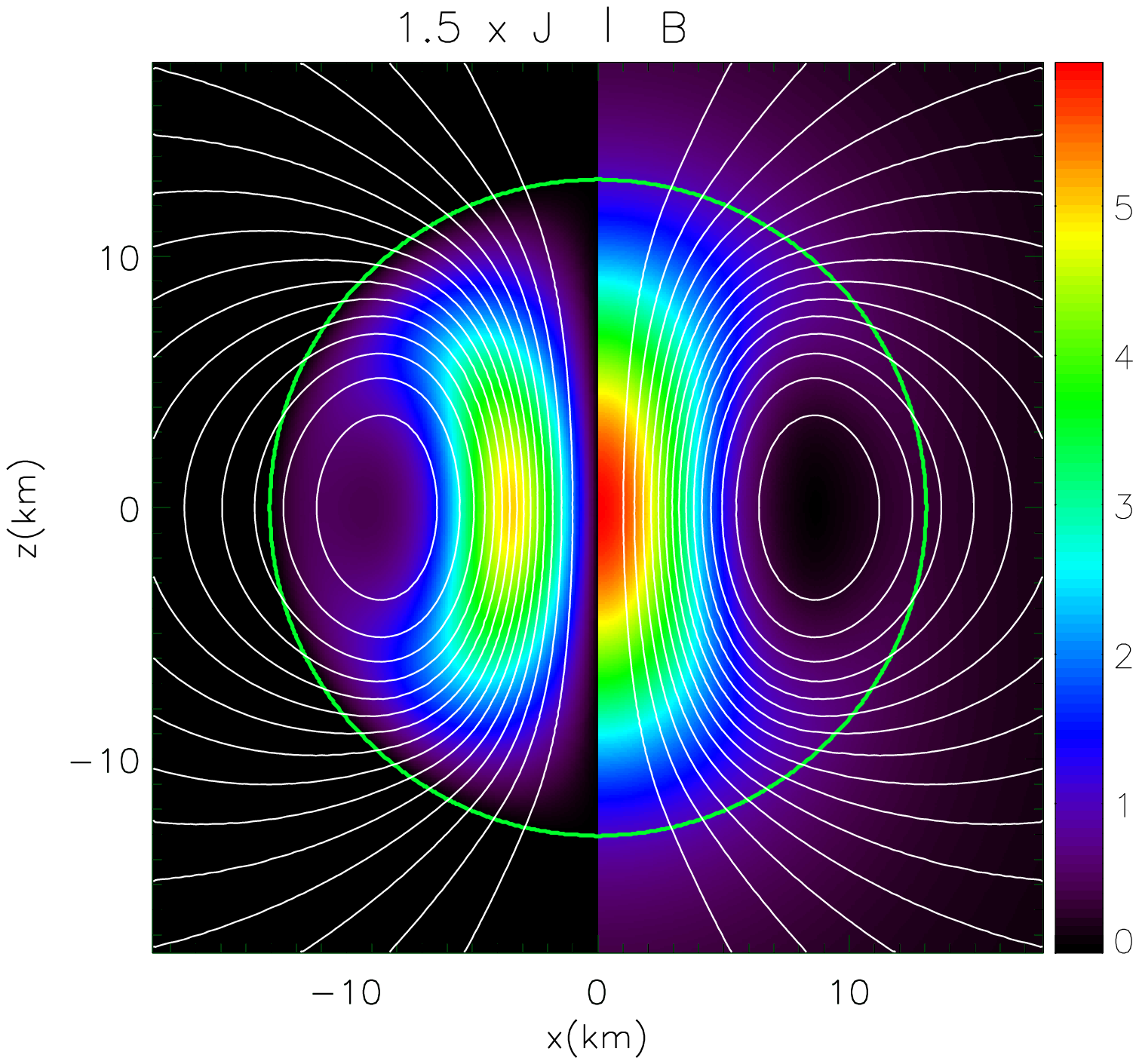}
	\includegraphics[width=.33\textwidth, bb=0 20 452 440, clip]{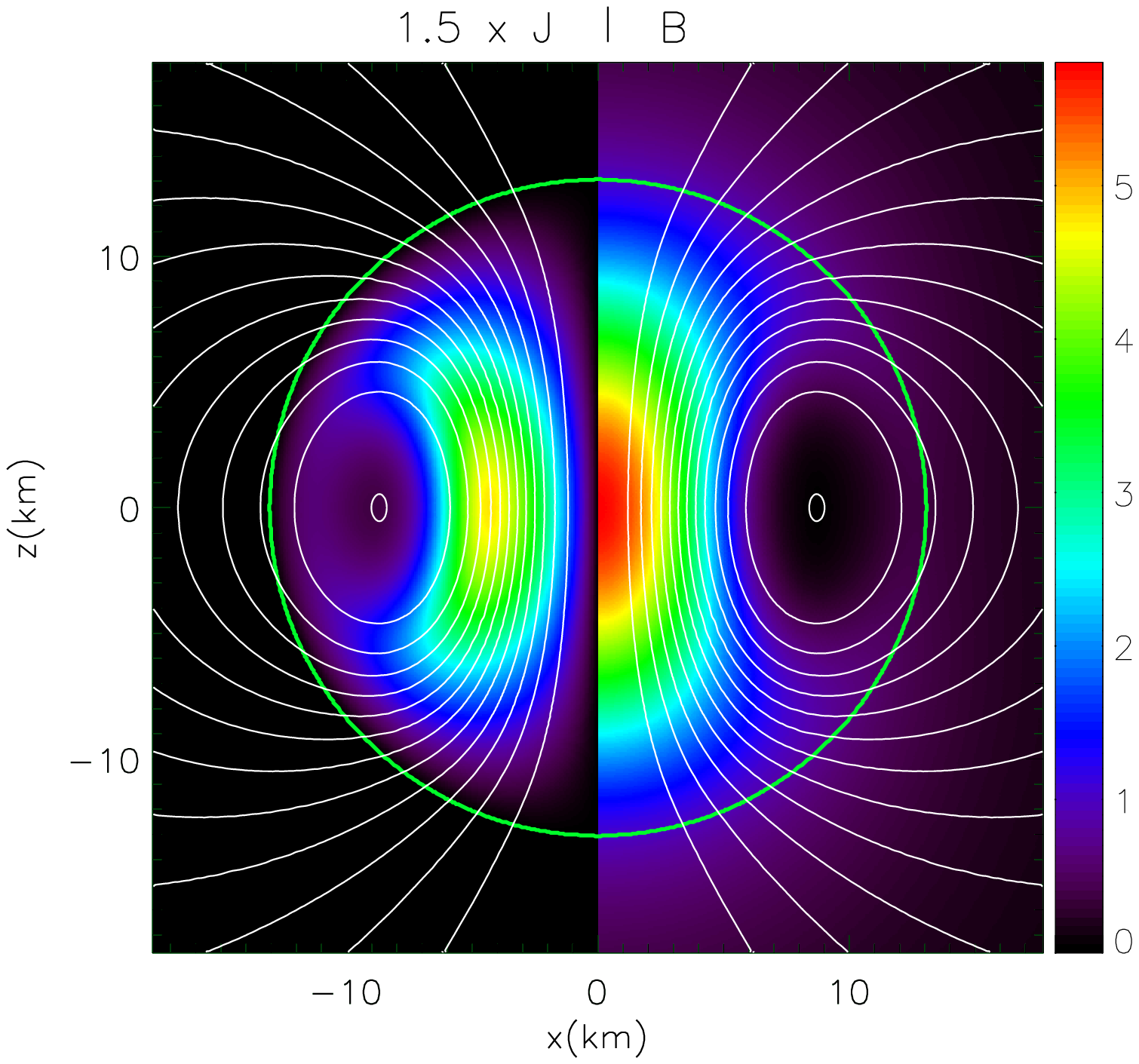}\\
    \includegraphics[width=.33\textwidth, bb=0 20 452 440, clip]{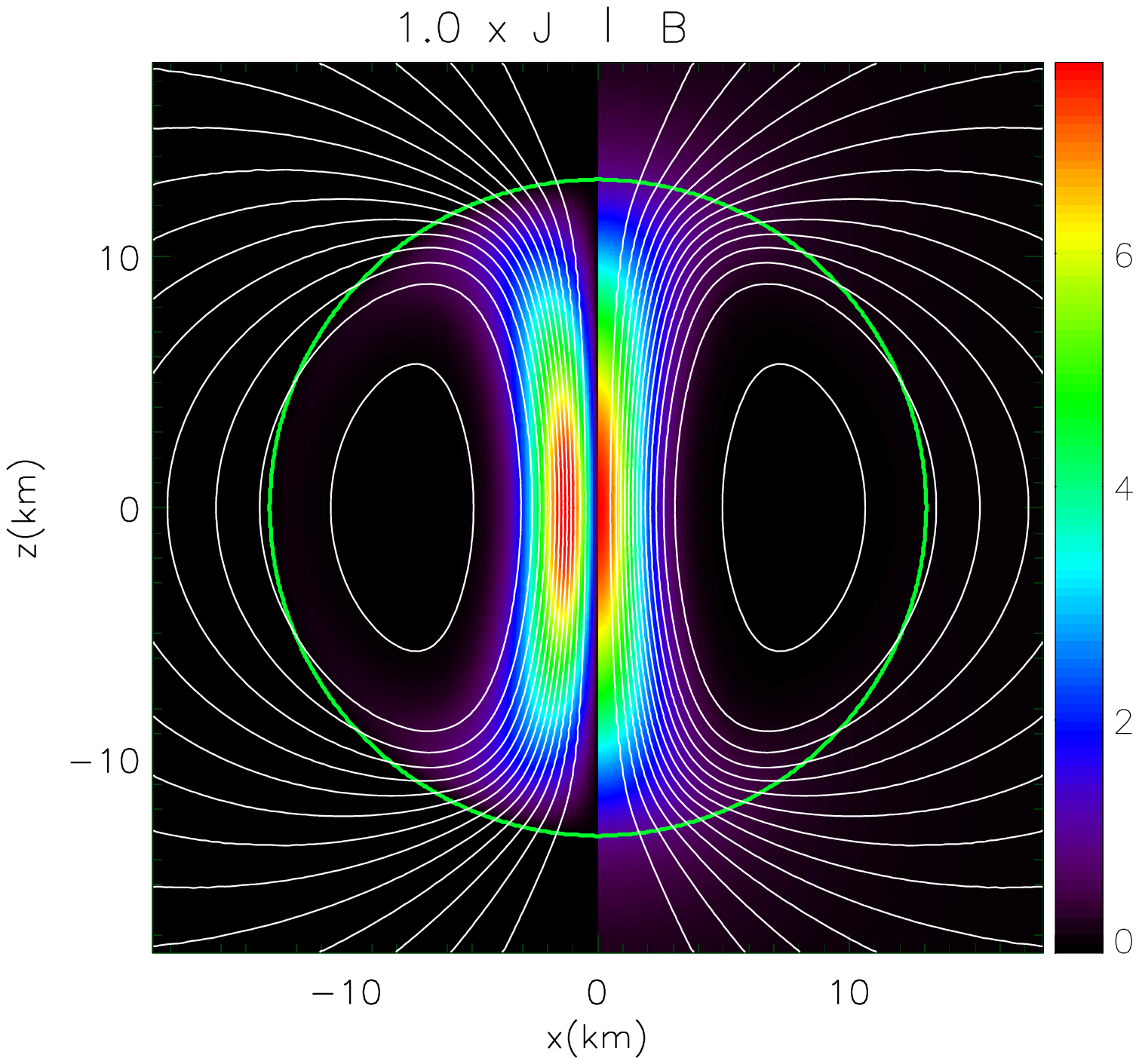}
	\includegraphics[width=.33\textwidth, bb=0 20 452 440, clip]{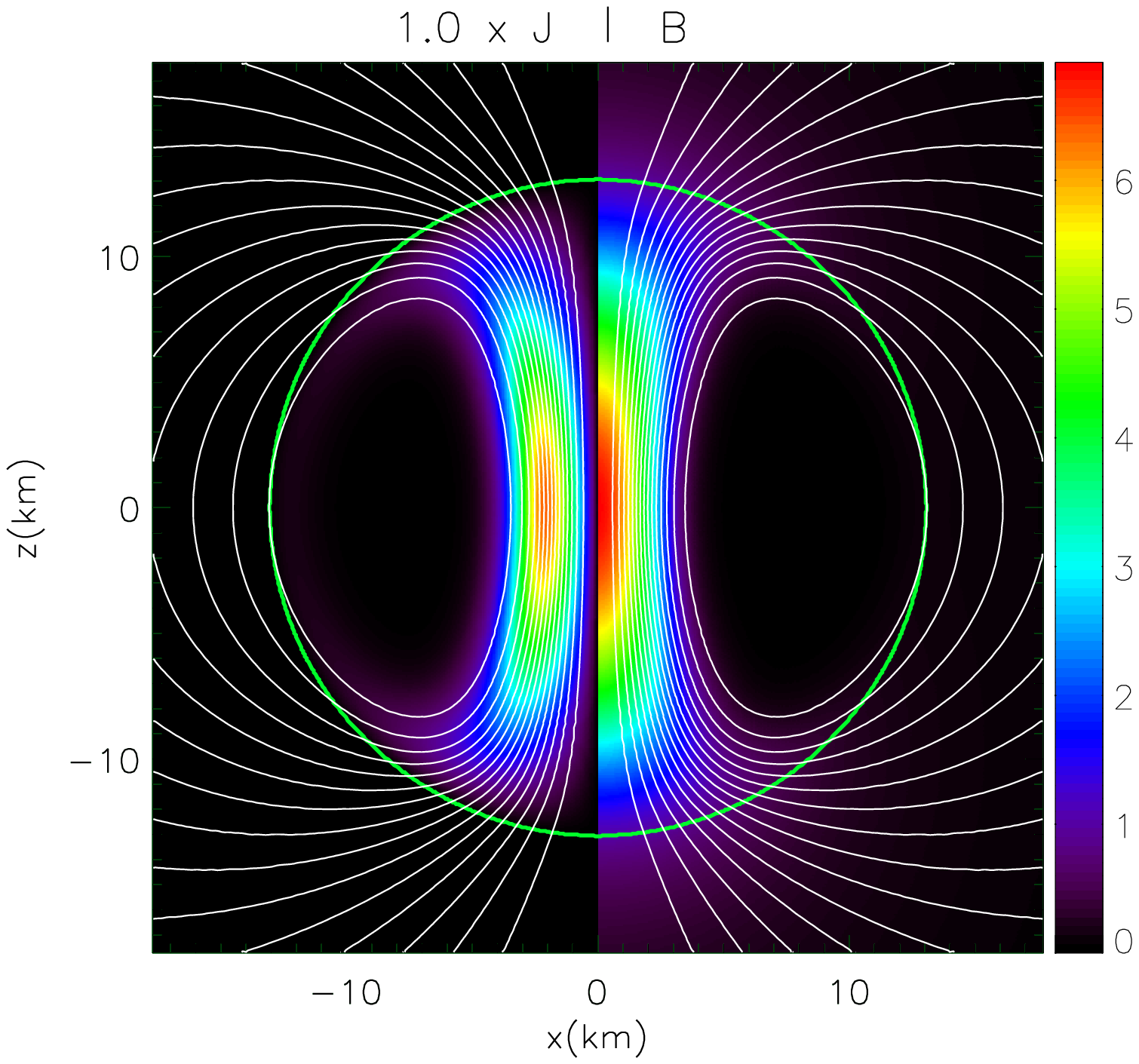}
	\includegraphics[width=.33\textwidth, bb=0 20 452 440, clip]{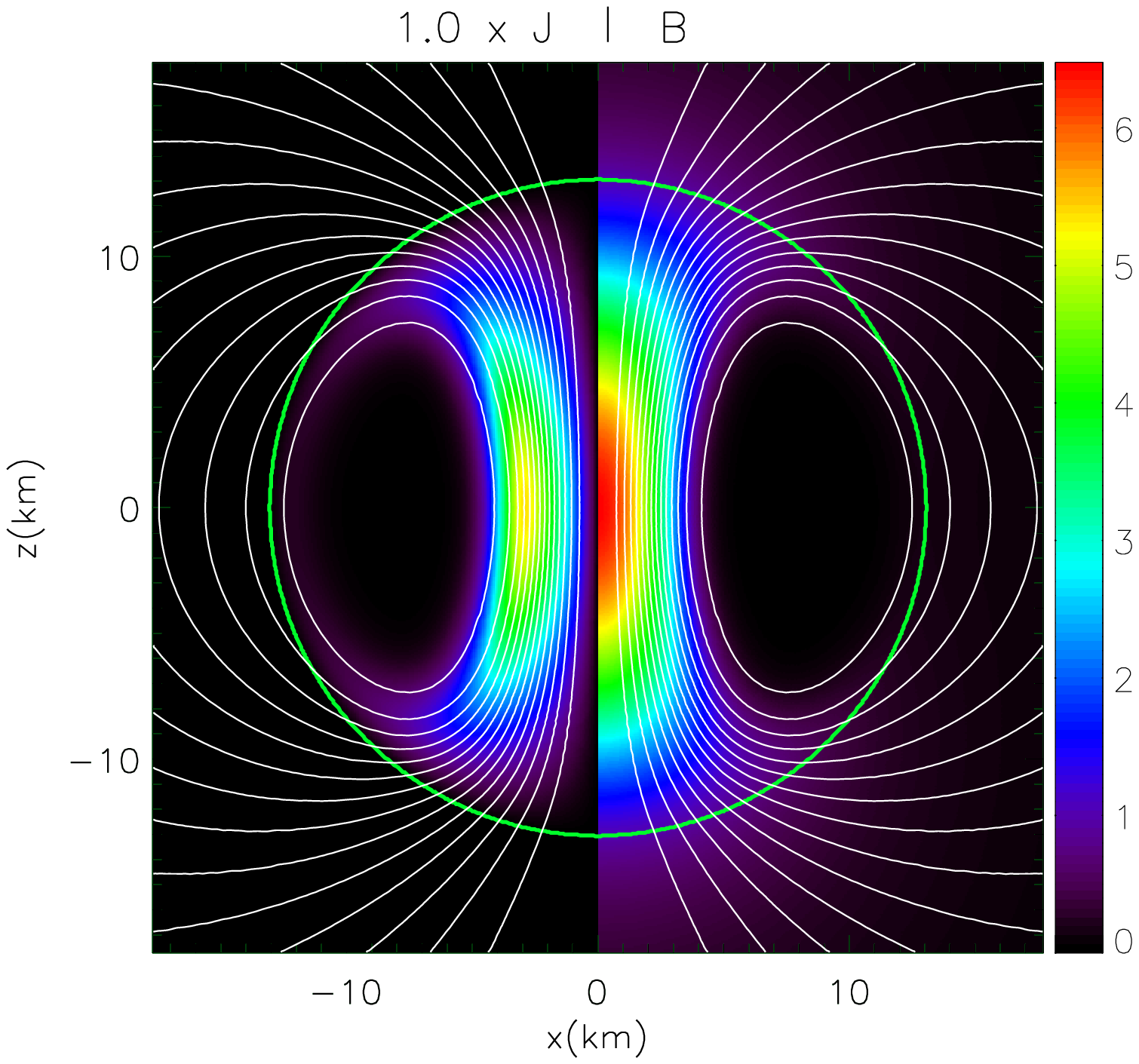}
	\caption{Purely poloidal field case. Strength of the azimuthal current in units of
          $10^{19}$G s$^{-1}$ (left half of each panel) and strength of
          the poloidal magnetic field in units $10^{14}$ G  (right half of each panel). White contours represent 
			magnetic field surfaces (isocontours of
                        $A_\phi$). The left column represents cases
                        with $\nu=1$, the central one those with $\nu=4$,
                        the right one those with $\nu=10$. From top to
                        bottom, rows represent cases with
                        $\xi=-0.5,-0.9,-1.0$.
                        The thick green line is the stellar
                        surface. In all cases the surface magnetic
                        field at the pole is $10^{14}$ G. Axes refer
                        to a cartesian frame centered on
                        the origin and with the $z$-axis corresponding
                        to the symmetry axis.
              }
	\label{fig:poloidal1}
\end{figure*}

In Fig.~\ref{fig:poloidal1} we show the magnetic field and the current
distribution for a series of models computed with different values of
$\xi < 0$ and different values for the poloidal index $\nu$. The effect of
the nonlinear term is to suppress the currents in the outer part of
the star, and to concentrate them in the inner region. The same holds
for the magnetic field. As $\xi$ decreases, the interior of the star becomes
progressively less magnetized, and the magnetic field is confined toward
the axis. It is interesting to note that this effect becomes
significative only as $\xi$ approaches $-1.0$ (for values of
$\xi$ closer to $0$ deviations are marginal). Moreover it is evident that in the
case of subtractive currents the magnetic field geometry that one finds
is almost independent on the magnetization index $\nu$. Indeed the
change in poloidal index  seems only to produce marginal effects in the magnetic
field distribution, with configurations that are slightly more concentrated
toward the axis for smaller values of $\nu$. In particular we find
that the unmagnetized and current free region extends to fill the
outer half of the star (the magnetic field at the equator drops to
zero at about half the stellar radius).  One also finds, in general,
that the ratio of the strength of the magnetic field at the pole, with
respect to the one at the centre increases by about 30 to 50\%, as
$\xi$ approaches $-1$.

Interestingly, we were not able
to obtain models with $\xi<-1$. This implies that we cannot find
configurations where there is a current inversion (the sign of the current is
always the same inside the star). Our relaxation scheme for
the GS equation seems at first to converge to a metastable equilibrium with
accuracy $\sim 10^{-4}$, but then the solution diverges. We want to
stress here that the Grad-Shafranov equation, in cases where the currents are nonlinear in the vector potential
$A_\phi$, becomes a nonlinear Poisson-like equation, that in
principle might admit multiple solutions and bifurcations (local
uniqueness is not guaranteed). This is a known problem \citep{Ilgisonis_Pozdnyakov03a}, and
suggests that a very small tolerance (we adopt $10^{-8}$) is required
to safely accept the convergence of a solution. This issue might be
related to the problem of local uniqueness for nonlinear elliptical
equations.  It is well known that the nonlinear Poisson equations of
the kind $\nabla^2\psi = k\psi^a$ satify local uniqueness only if $ka
\ge 0$. It is evident that this depends on the relative sign of the coefficient and
exponent of the nonliner source term: in our case the relative sign of
$\xi$ and $\nu$. Given that $\nu$ is always positive, what matters is
just the sign of $\xi$. This explains why we can obtain solutions with
additive currents ($\xi >0$) even in the regime dominated by the
nonlinear term, while solutions with subtractive currents ($\xi <0$) can only be
built up to $\xi >-1$, where the contribution of the nonlinear current is
still smaller than the linear one which act as a stabilizing
term. However we want to recall here that the Grad-Shafranov is not a Poisson equation, and it is
not proved that the same uniqueness criteria apply.

\begin{figure*}
	\centering
	\includegraphics[width=.33\textwidth, bb=0 20 452 440, clip]{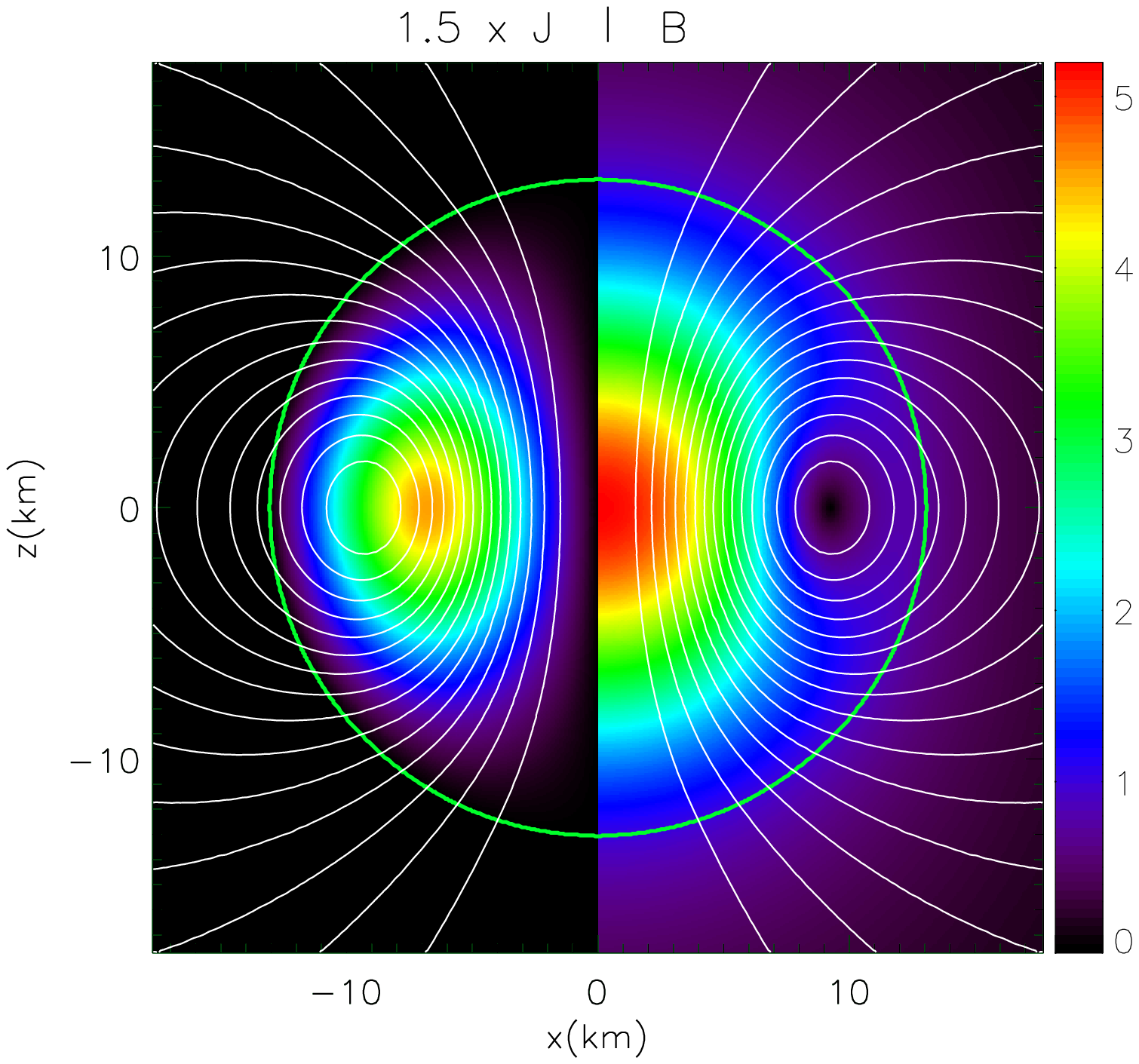}
	\includegraphics[width=.33\textwidth, bb=0 20 452 440, clip]{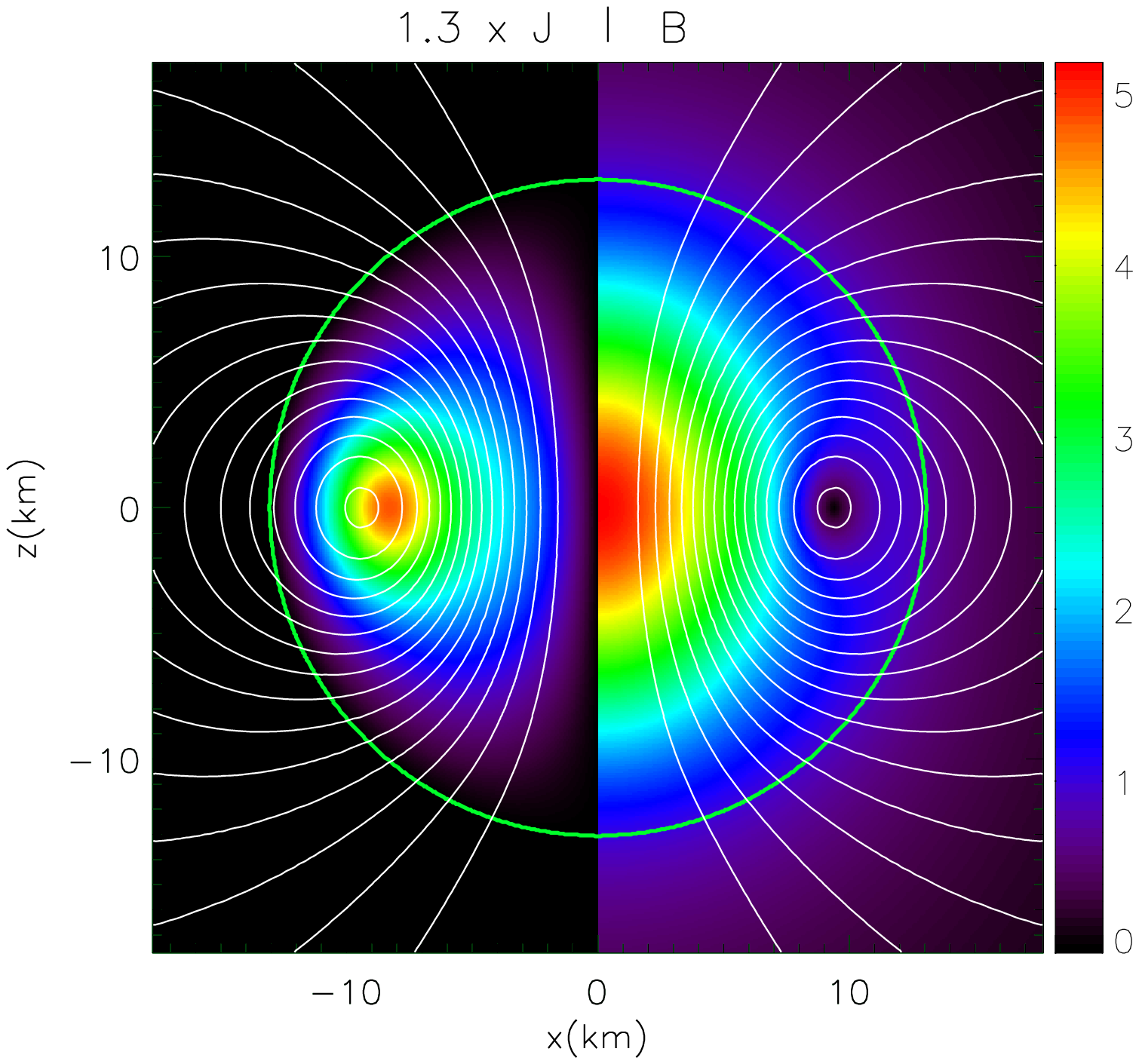}
	\includegraphics[width=.33\textwidth, bb=0 20 452 440, clip]{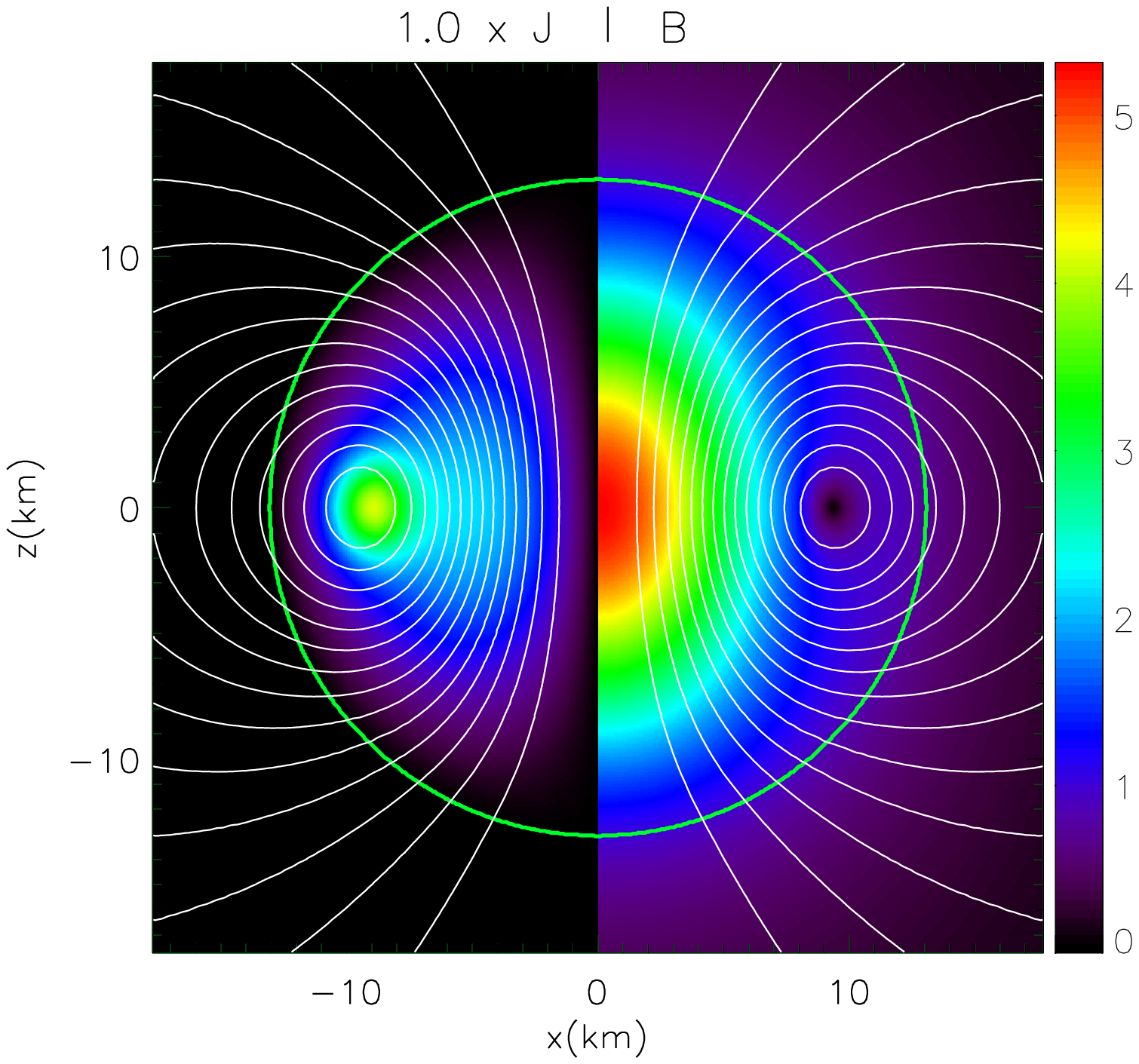}\\
    \includegraphics[width=.33\textwidth, bb=0 20 452 440, clip]{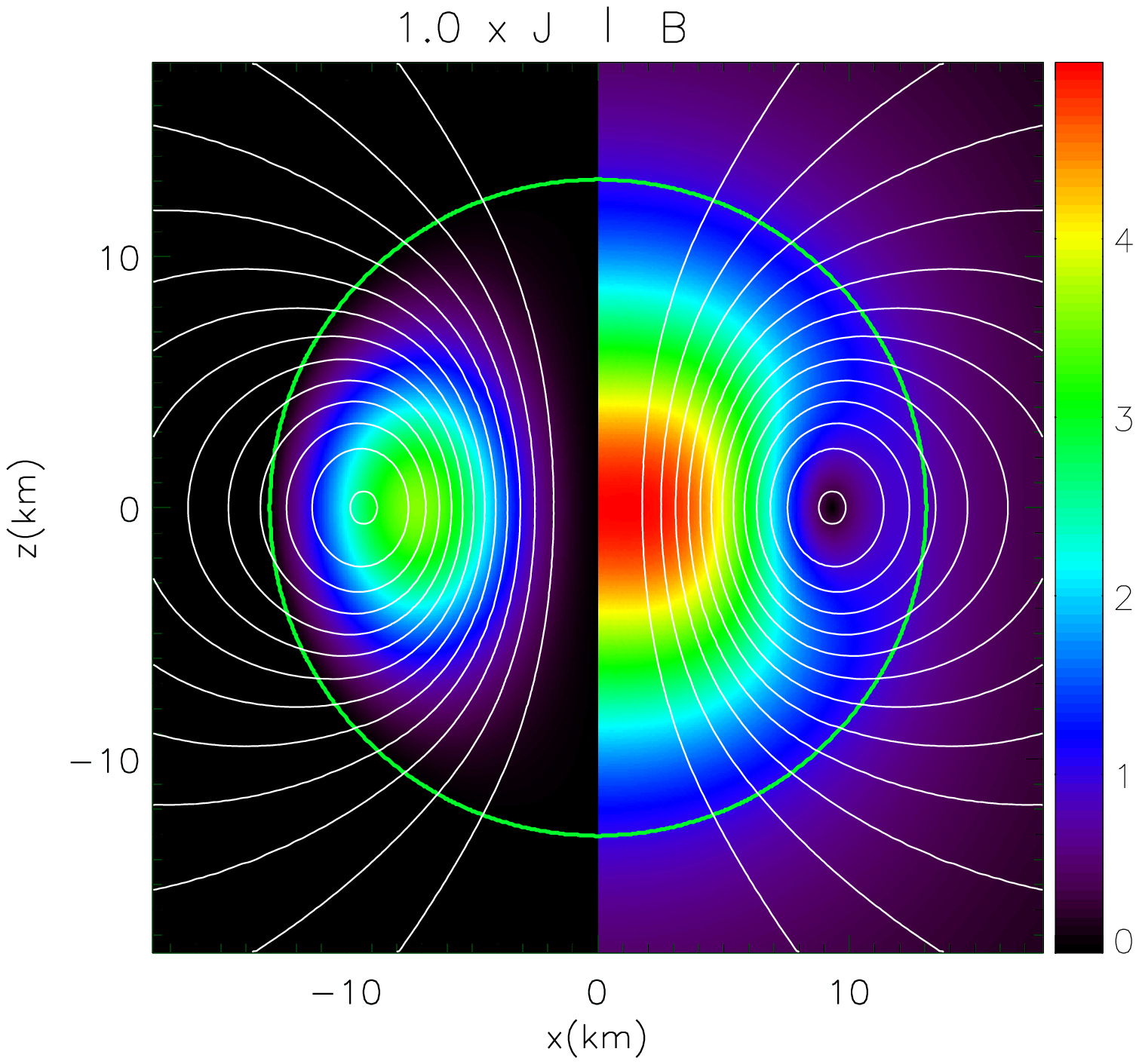}
	\includegraphics[width=.33\textwidth, bb=0 20 452 440, clip]{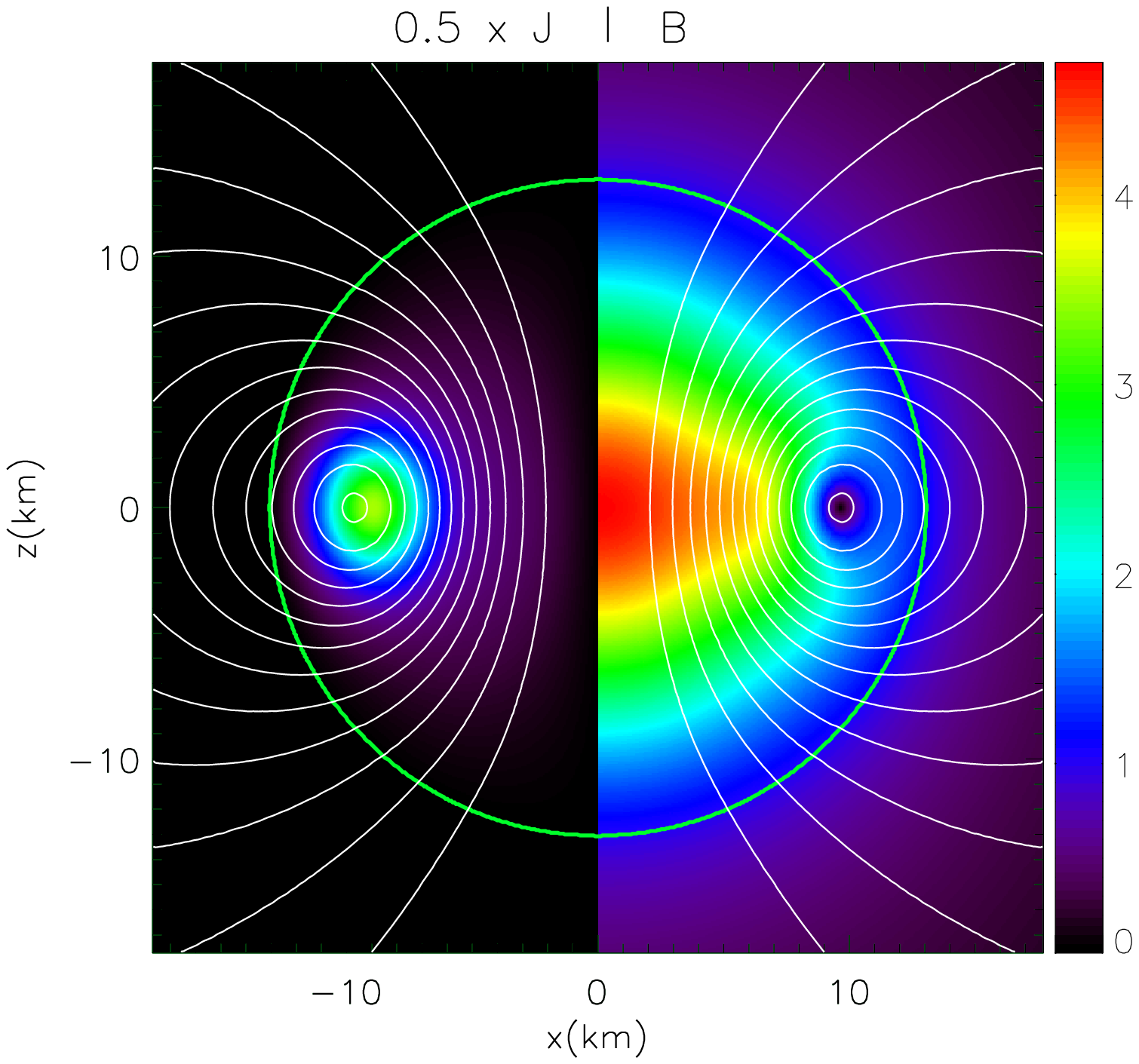}
	\includegraphics[width=.33\textwidth, bb=0 20 452 440, clip]{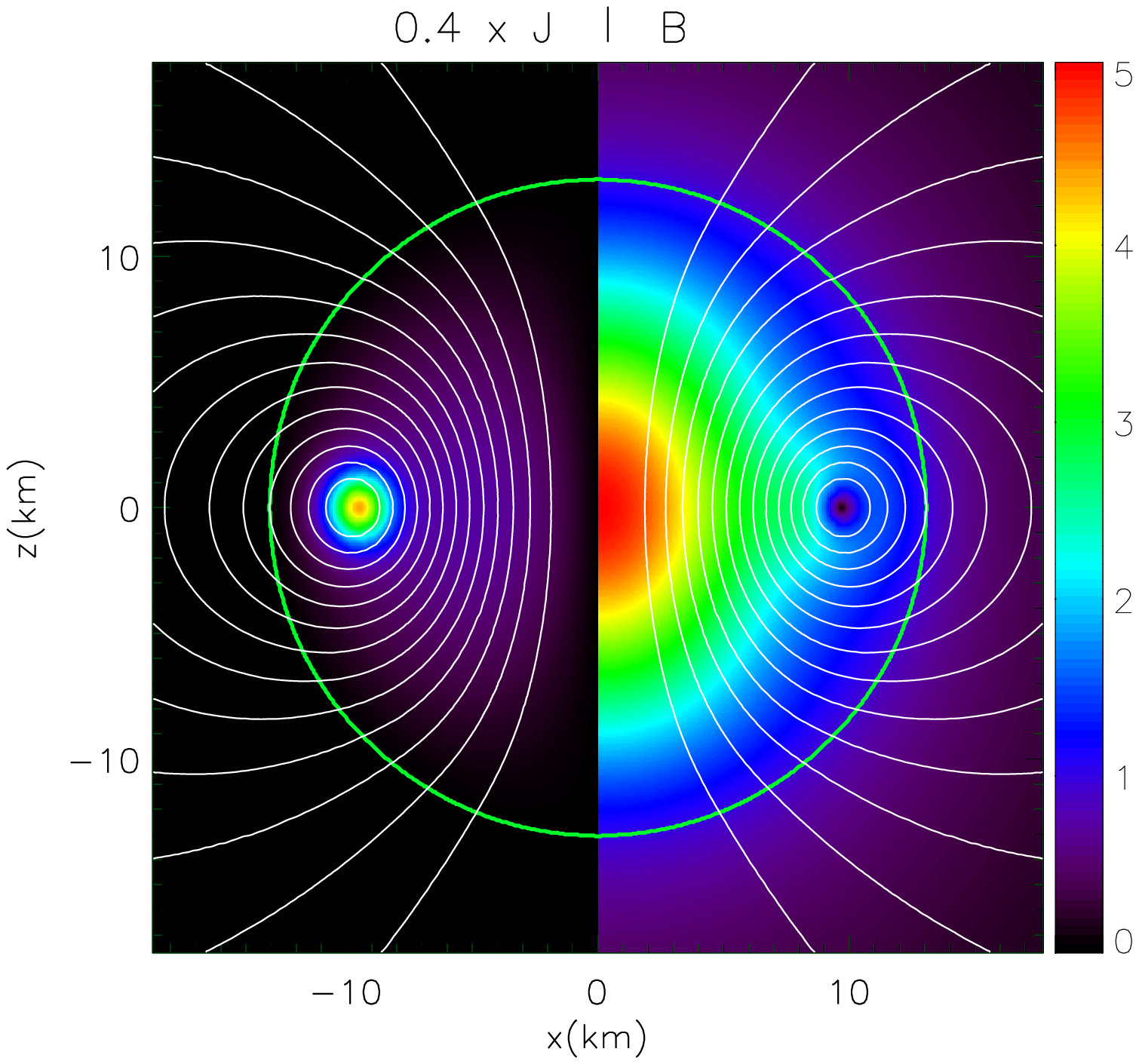}\\
    \includegraphics[width=.33\textwidth, bb=0 20 452 440, clip]{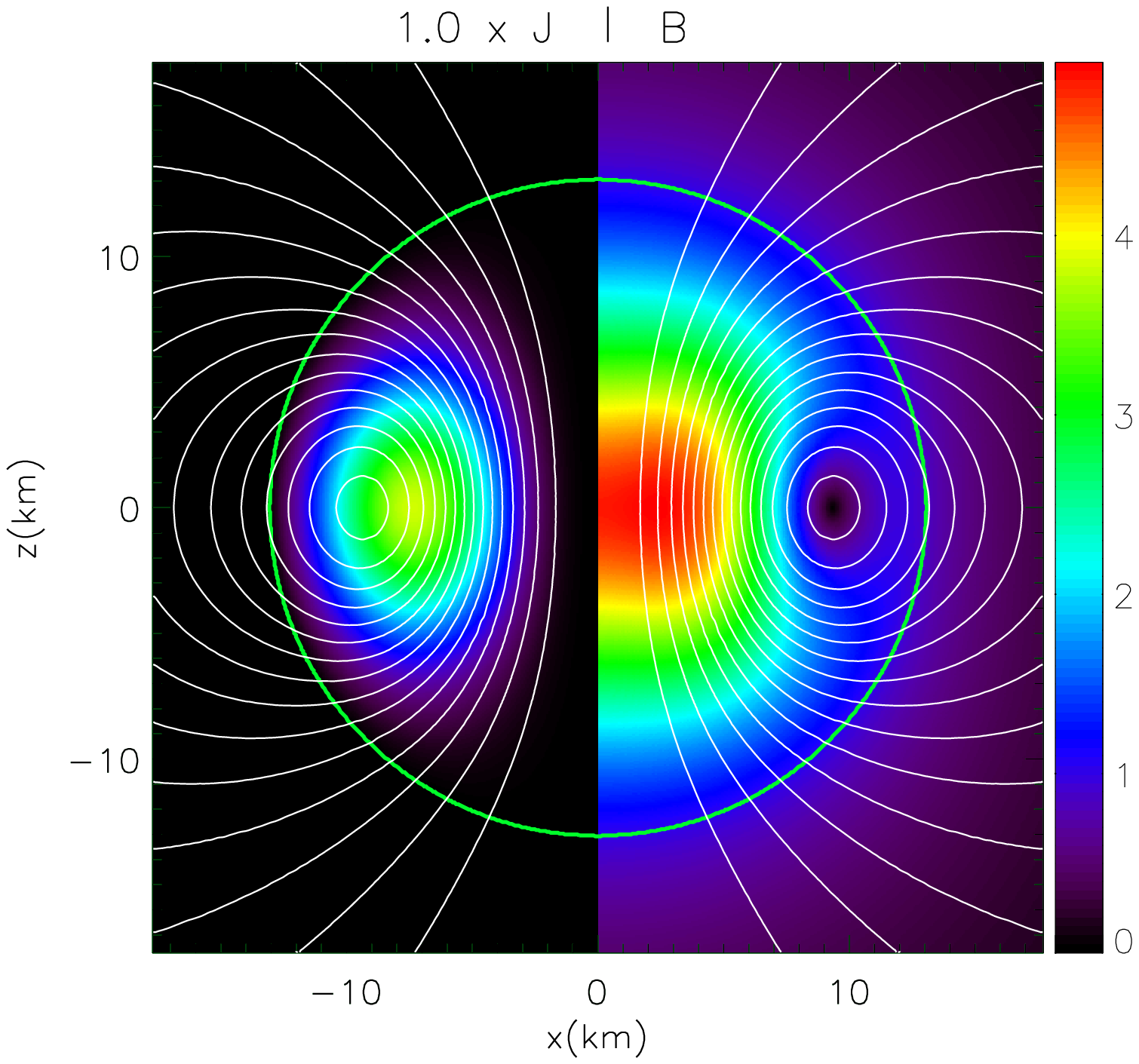}
	\includegraphics[width=.33\textwidth, bb=0 20 452 440, clip]{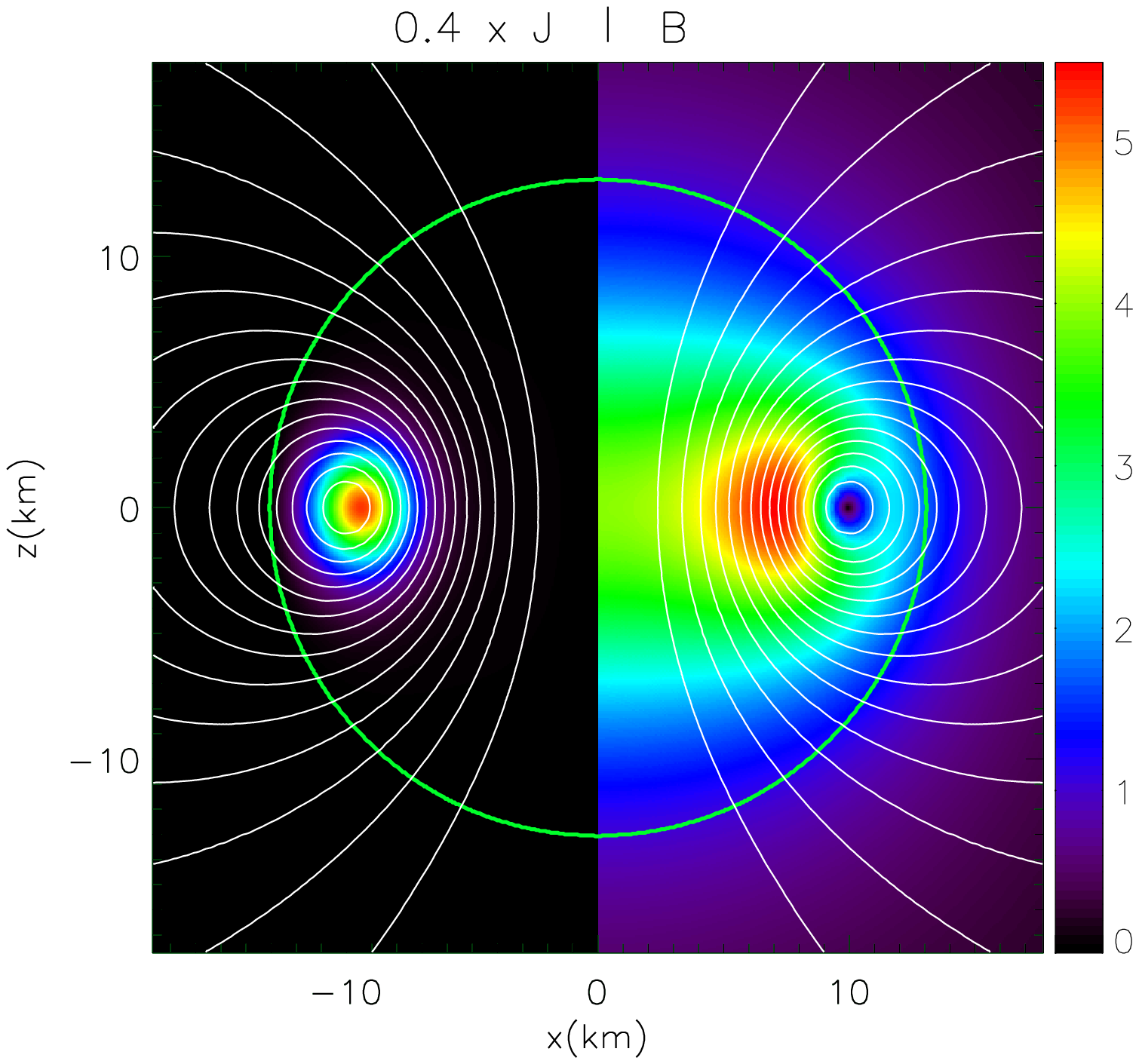}
	\includegraphics[width=.33\textwidth, bb=0 20 452 440, clip]{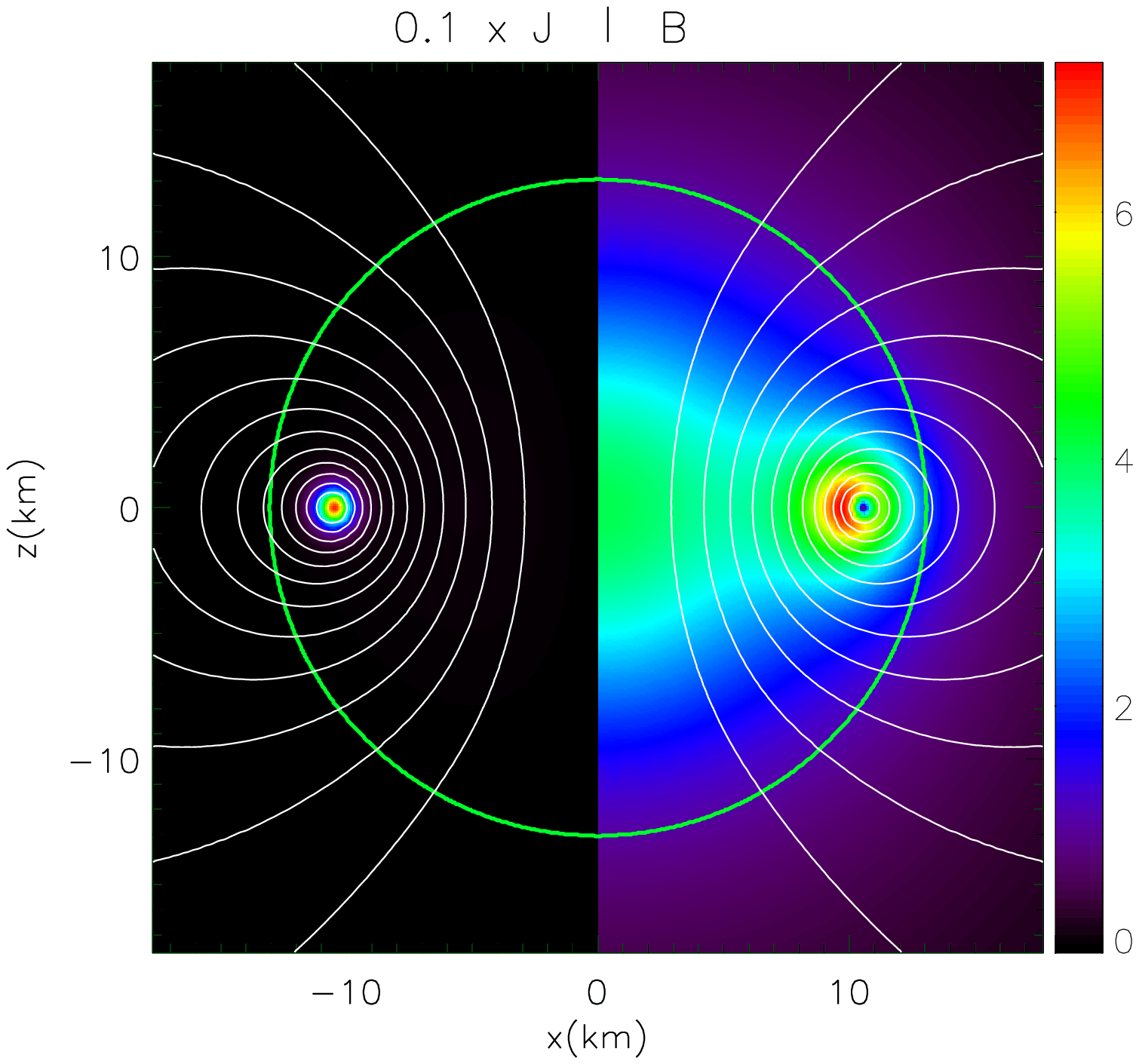}
	\caption{Purely poloidal field case. Strength of the azimuthal current in units of
          $10^{19}$G s$^{-1}$ (left half of each panel) and strength of
          the poloidal magnetic field in units $10^{14}$ G  (right half of each panel). White contours represent 
			magnetic field surfaces (isocontours of
                        $A_\phi$). The left column represents cases
                        with $\nu=1$, the central one those with $\nu=4$,
                        the right one those with $\nu=10$. From top to
                        bottom, rows represent cases with
                        $\xi=2.0,10.0,200.0$.
                        The thick green line is the stellar
                        surface. In all cases the surface magnetic
                        field at the pole is $10^{14}$ G. Axes refer
                        to a cartesian frame centered on
                        the origin and with the $z$-axis corresponding
                        to the symmetry axis.
              }
	\label{fig:poloidal2}
\end{figure*}

In Fig.~\ref{fig:poloidal2} we show the opposite case of additive
currents, $\xi>0$. The value of $\nu$ in this case establishes how much
concentrated these currents are, and plays a major role in determining
the properties of the resulting magnetic field. Rising the value of
$\xi$ the nonlinear
 currents become progressively more important. We can define
a nonlinear dominated regime in the limit of 
high $\xi$, where the magnetic field structure and distribution
 converge to a solution that is independent of $\xi$. The values of $\xi$, at which this limit is reached,
depends on $\nu$. For $\nu=1$ the limit is achieved already at $\xi
=20$ as can be inferred from Fig.~\ref{fig:poloidal2}, while for $\nu=10$
the limit is reached at $\xi \sim 1000$. As already noted \citep{Pili_Bucciantini+14a},
in the case $\nu=1$ the presence of a nonlinear current term does not
alter significantly the geometry of the magnetic field, or other
global integrated quantities like the net global dipole moment. The ratio of
the strength of the magnetic field at the centre with respect to the
one at the pole diminishes slightly by about 10\%. The location of the
neutral current point, where the magnetic field vanishes is unchanged.

At higher values of $\nu$ the magnetic field geometry in the
nonlinear dominated regime changes substantially. The overall current is
strongly concentrated around the neutral point. The location of the
neutral point itself shifts toward the surface of the NS, from about 0.7
stellar radii at $\xi=0$ to about 0.8 stellar radii in the nonlinear
dominated limit. Moreover the maximum in the strength of the magnetic
field is not reached at the centre any longer, but at intermediate radii
where the nonlinear current is located. In this case the value of this local
maximum can be a factor a few higher than the value at the centre. Configurations with two local
maxima are also possible. This behavior is strongly reminiscent of
what is found for the so-called TT configurations, where a
toroidal component of the magnetic field is also present, inducing a
current that behaves as the nonlinear term we have introduced
here (see next section).

\begin{figure*}
	\centering
	\includegraphics[width=.33\textwidth, bb=0 0 452 440, clip]{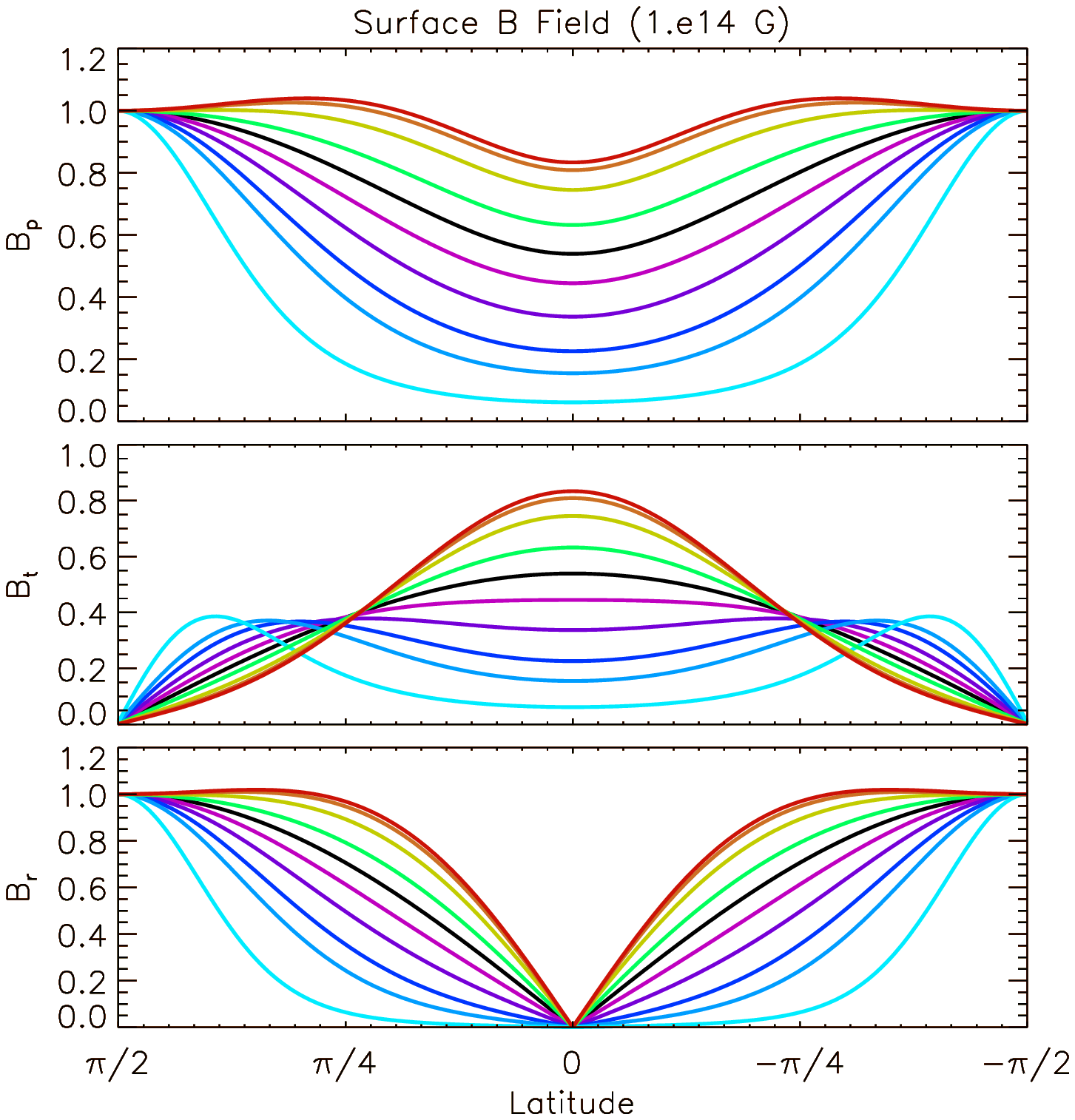}
	\includegraphics[width=.33\textwidth, bb=0 0 452 440, clip]{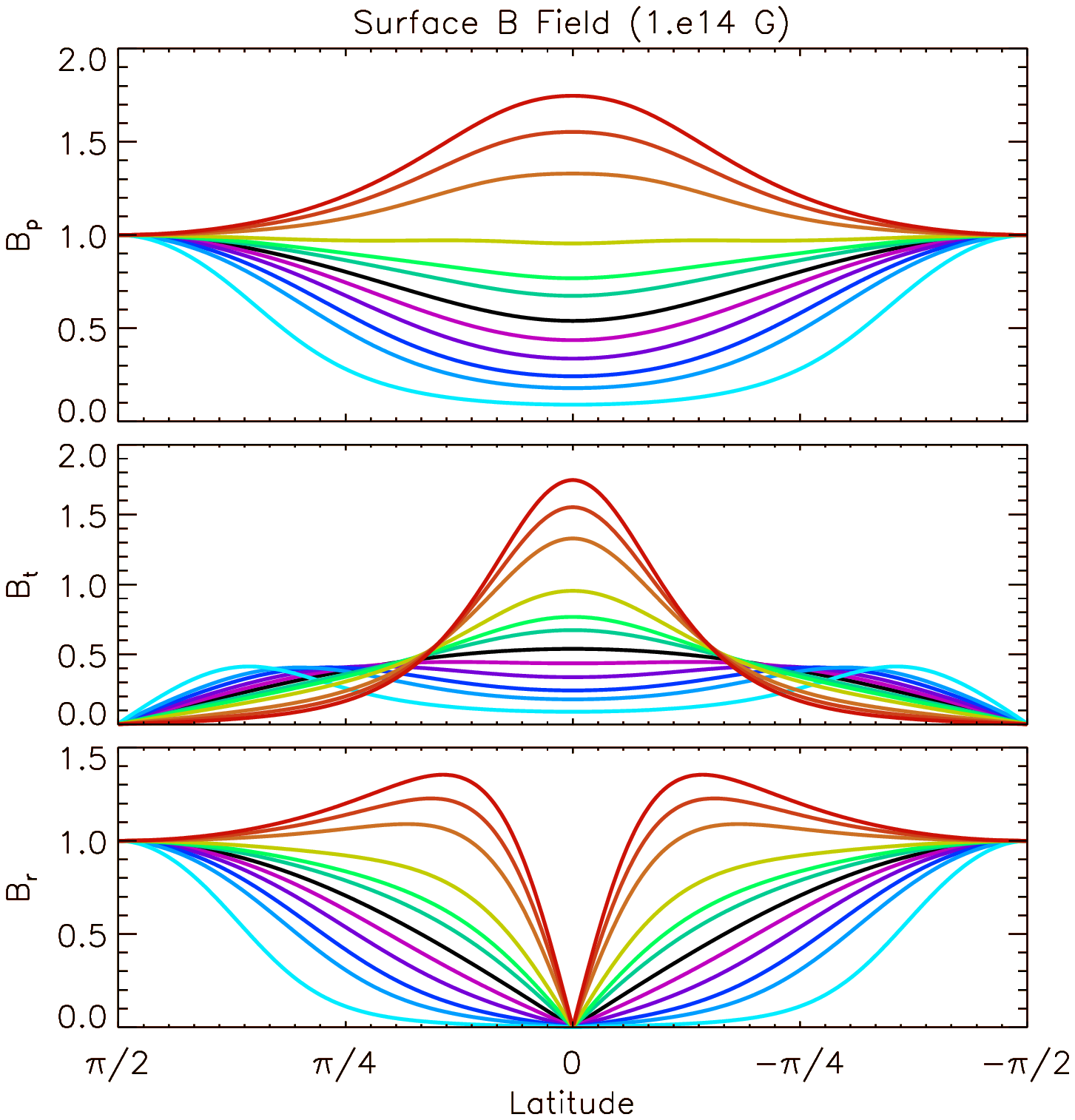}
	\includegraphics[width=.33\textwidth, bb=0 0 452 440,  clip]{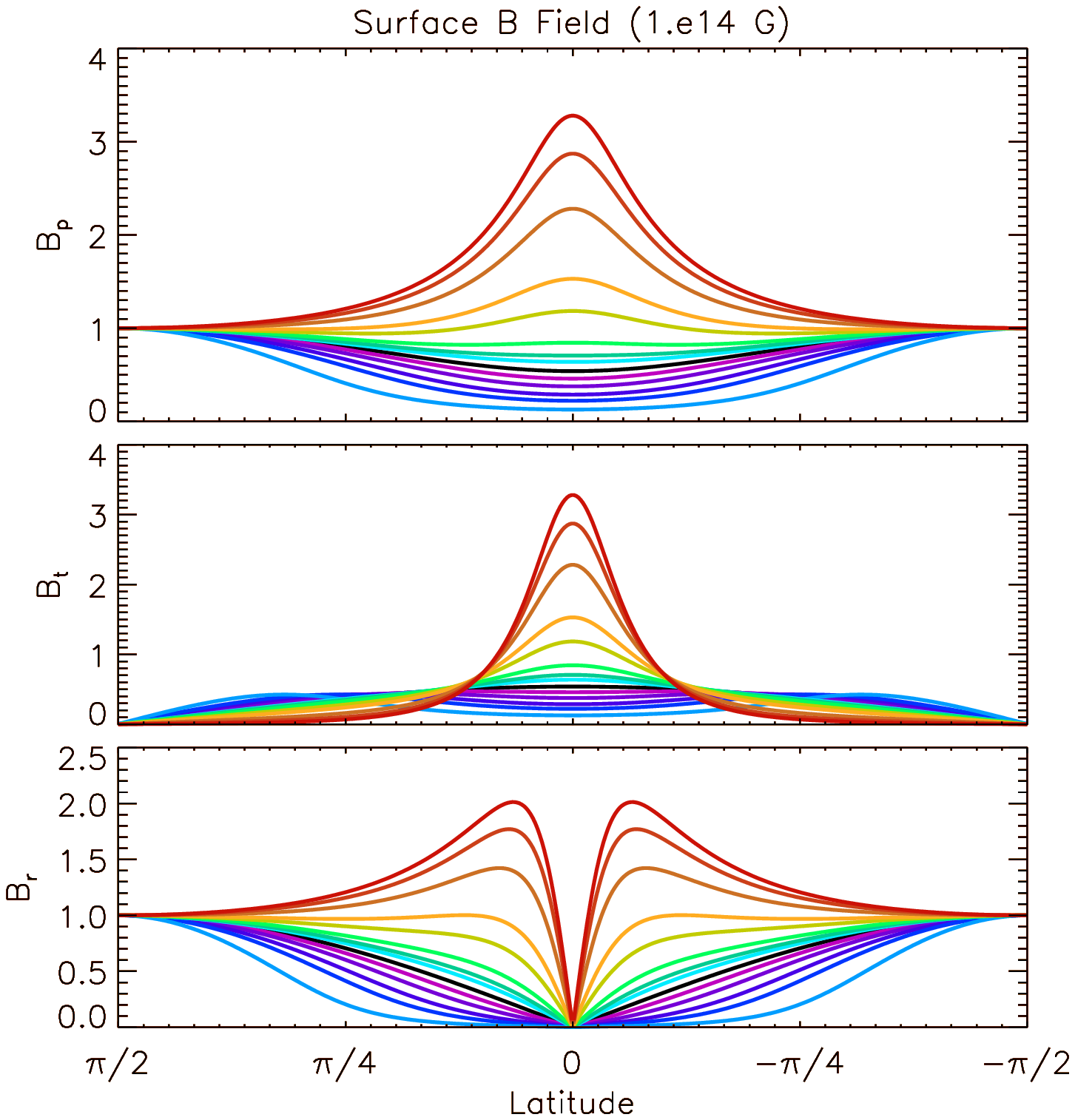}\\
	\includegraphics[width=.33\textwidth, bb=0 0 452 60, clip]{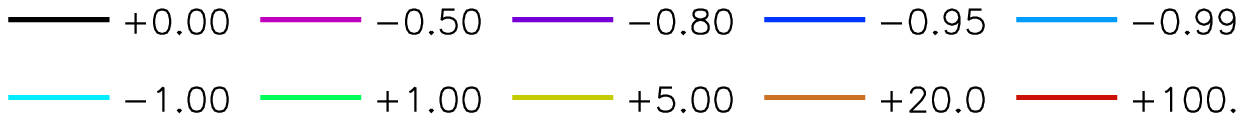}
	\includegraphics[width=.33\textwidth, bb=0 0 452 60, clip]{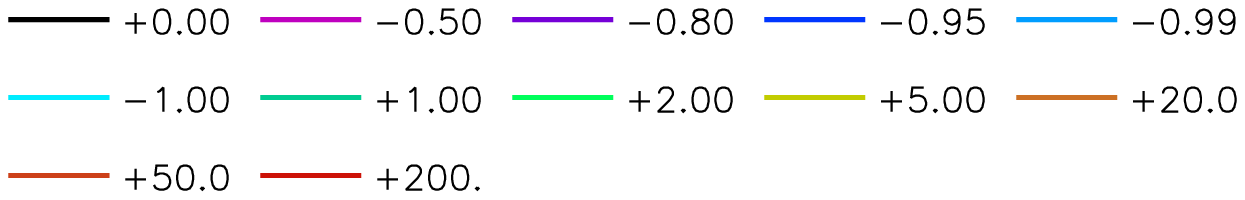}
	\includegraphics[width=.33\textwidth, bb=0 0 452 60, clip]{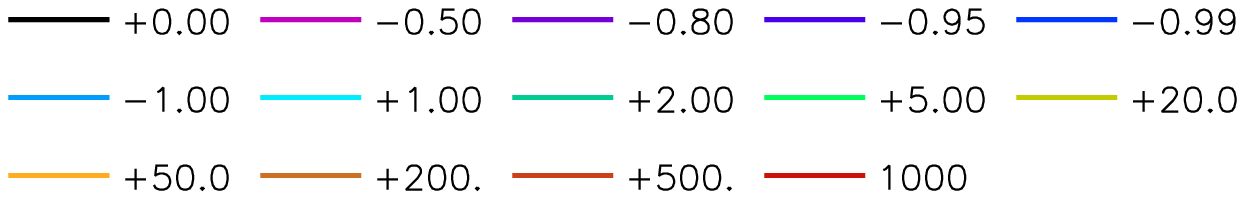}\\
\caption{Purely poloidal field case. Magnetic field at the surface normalized to the value at the
  pole, for various values of $\xi$. Left column represents cases with $\nu=1$, central column
  cases with $\nu=4$ and right column cases with $\nu=10$. Upper
  panels display the total strength of the poloidal magnetic field,
  middle panels the strength of the parallel $\theta$ component, and
  lower panels the radial one.
}
\label{fig:polsurface}
\end{figure*}

One can also look at the strength and distribution of the surface
magnetic field, shown in Fig.~\ref{fig:polsurface}. For
decreasing values of $\xi <0$ the magnetic field tends to concentrate at
the pole, in a region that is $\sim 30^\circ$ for $\xi=-1$. The radial
magnetic field in the equatorial region is strongly suppressed, the
field is almost parallel to the stellar surface, and the
overall strength of the poloidal field is a factor
10 smaller with respect to the case with $\xi=0$ . In general these results are 
weakly dependent on the value of $\nu$, with higher values of $\nu$
leading to configurations where the field is slightly less concentrated toward
the poles. A quite different behavior is seen for the cases of additive
currents $\xi >0$. For $\nu=1$, the radial component of the magnetic field
tends to be higher than in the case $\xi=0$, and it tends to be
uniform in the polar region. The $\theta$ component of the magnetic
field increases in the equatorial region by about a factor 2. The
overall strength of the magnetic field becomes quite uniform over the
stellar surface in the nonlinear dominated regime. These effects are
further enhanced for increasing values of $\nu$. At $\nu=4$, in the
nonlinear dominated regime, the radial component of the magnetic field
reaches its maximum at $\sim \pm 25^\circ$ from the equator. The
$\theta$ component, parallel to the NS surface, is instead strongly
enhanced by about a factor 3 at the equator. The result is that for
increasing $\xi$ there is a transition from configurations where the
poloidal field strength is higher at the poles, to configurations where
it is higher (by about 40\%) at the equator, with intermediate cases
where it can be almost uniform.  At $\nu=10$ these effects are even
stronger: the radial field now peaks very close to the equator, at
$\sim \pm 10^\circ$, and the overall strength of the magnetic field can
be higher at the equator by a factor $\sim 3$ with respect to the
poles. This is the clear manifestation of a concentrated and
localized peripheral current, close to the surface of the star. 

To summarize the results in the fully saturated non-linear regime:
\begin{itemize}
\item subtractive currents, independent of their functional form,
  confine the magnetic field toward the axis, leaving large
  unmagnetized region inside the star;
\item for subtractive currents, the surface magnetic field is
  concentrated in a polar region of $\sim 30^\circ$ from the pole,
  while at lower latitudes ($\pm 40^\circ$ from the equator) it can
  be a factor 10 smaller than at the pole (to be compared with one half
  for pure dipole); 
\item additive currents tend to concentrate the field in the outer
  layer of the star, the effect being stronger for higher values of
  the non-linearity; the field strength reaches its maximum closer to
  the surface, while its strength at the center can be even more than a
  factor 2 smaller;
\item for additive current, the structure of the field at the equator
  can be qualitatively different than a dipole: higher at the equator
  than at the pole, even by a factor a few. A geometry similar to what
  is found in TT configurations.
\end{itemize} 

\subsection{Twisted Torus Configurations}
\label{sec:tt}

\begin{figure}
	\centering
	\includegraphics[width=.50\textwidth]{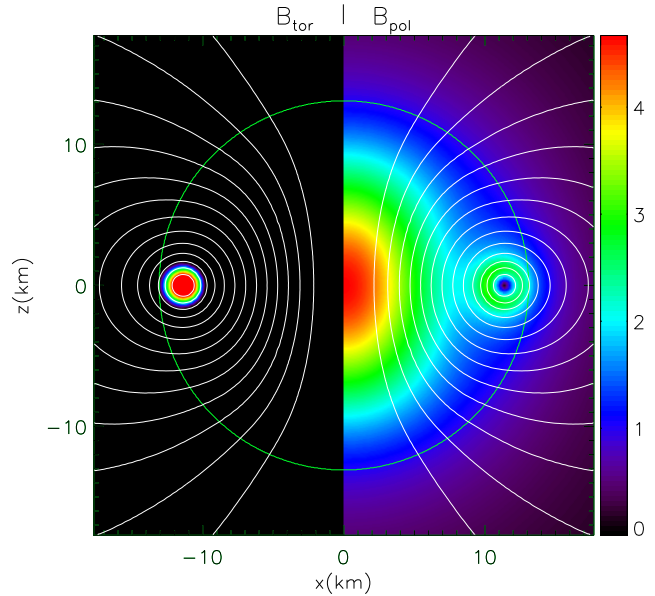}
\caption{Magnetic field for a TT configuration with
  $\zeta=0$ and $a=1.5$ (corresponding to the maximum of the ratio
  $\mathcal{H}_{\rm tor}/\mathcal{H}$). Strength of the toroidal magnetic field (left), and poloidal
  magnetic field (right) normalized to the surface value at the
  pole. White contours represent 
			magnetic field surfaces (isocontours of
                        $A_\phi$).  The thick green line is the stellar
                        surface. Axes refer
                        to a cartesian frame centered on
                        the origin and with the $z$-axis corresponding
                        to the symmetry axis.
}
\label{fig:tt}
\end{figure}

Mixed geometries with poloidal and toroidal magnetic fields have
been presented in the past in the so-called TT
  configurations
  \citep{Ciolfi_ferrari+09a,Ciolfi_Ferrari+10a,Ciolfi_Rezzolla13a,Lander_Jones09a,Glampedakis_Andersson+12a,Fujisawa_Yoshida+12a,Pili_Bucciantini+14a,Pili_Bucciantini+14b}. These
  configurations are characterized by a torus-like region, in the
  interior of the star, just
 under the stellar surface, where the toroidal field is
confined. This geometry can be obtained if one chooses for the current function
$\mathcal{I}$ the form of Eq.~\eqref{eq:fbern1}.  In Fig.\ref{fig:tt} we show the magnetic field
distribution for a typical TT solution.

Particular attention has been recently devoted to the study of this
kind of systems, because there is evidence that magnetic field, in a
fluid star, tends to
relax toward a TT geometry, and that only mixed
configurations can be dynamically stable
\citep{Braithwaite09a,Braithwaite_Norlund06a,Braithwaite_Spruit06a}. Motivated
by these dynamical studies, efforts in the past have gone
toward modeling systems where the equilibrium magnetic geometry was
such that the magnetic energy was dominated by
the toroidal component. Despite several attempts in various regimes
\citep{Ciolfi_ferrari+09a,Lander_Jones09a,Pili_Bucciantini+14a}, only configurations where the energetics was dominated by the
poloidal component could be found. Recently \citet{Ciolfi_Rezzolla13a} (CR13 hereafter) have shown that a
very peculiar current distribution might be required in order to
obtain toroidally dominated systems. This raises questions about
the importance of the specific choice in the form of currents
$\mathcal{I}$ and $\mathcal{M}$. More precisely one would like to
know if previous failure to get toroidally dominated geometries is
due to a limited sample of the parameter space, or if only very \textit{ad hoc}
choices for the current distribution satisfy this requirement. 
Moreover most of the efforts have
concentrated onto understanding how this magnetic field acts on the star,
and the amount of deformation that it induces. This is mostly
motivated by searches for possible gravitational waves from neutron stars. Attention has focused on a
limited set of models, and current distributions. In particular a deep
investigation has been carried out only for the case $\zeta=0$ and $\zeta=0.1$
\citep{Lander_Jones09a,Pili_Bucciantini+14a}. 

Here we present a full investigation of TT configurations for various
values of the parameter $\zeta$. This parameter regulates the shape of
the current distribution inside the torus. For $\zeta \rightarrow -0.5$ the current
becomes uniformly distributed within the torus, while for $\zeta > 0$
it concentrates in the vicinity of the neutral line, where the
poloidal field vanishes. It was shown that it is the integrated
current associated with the current function $\mathcal{I}$ that
prevents TT configurations to reach the toroidal dominated regime. As
the strength of this current increases, the toroidal field rises, but
the torus-like region shrinks toward the surface of the star and its volume
diminishes. 

\begin{figure}
	\centering
	\includegraphics[width=.50\textwidth]{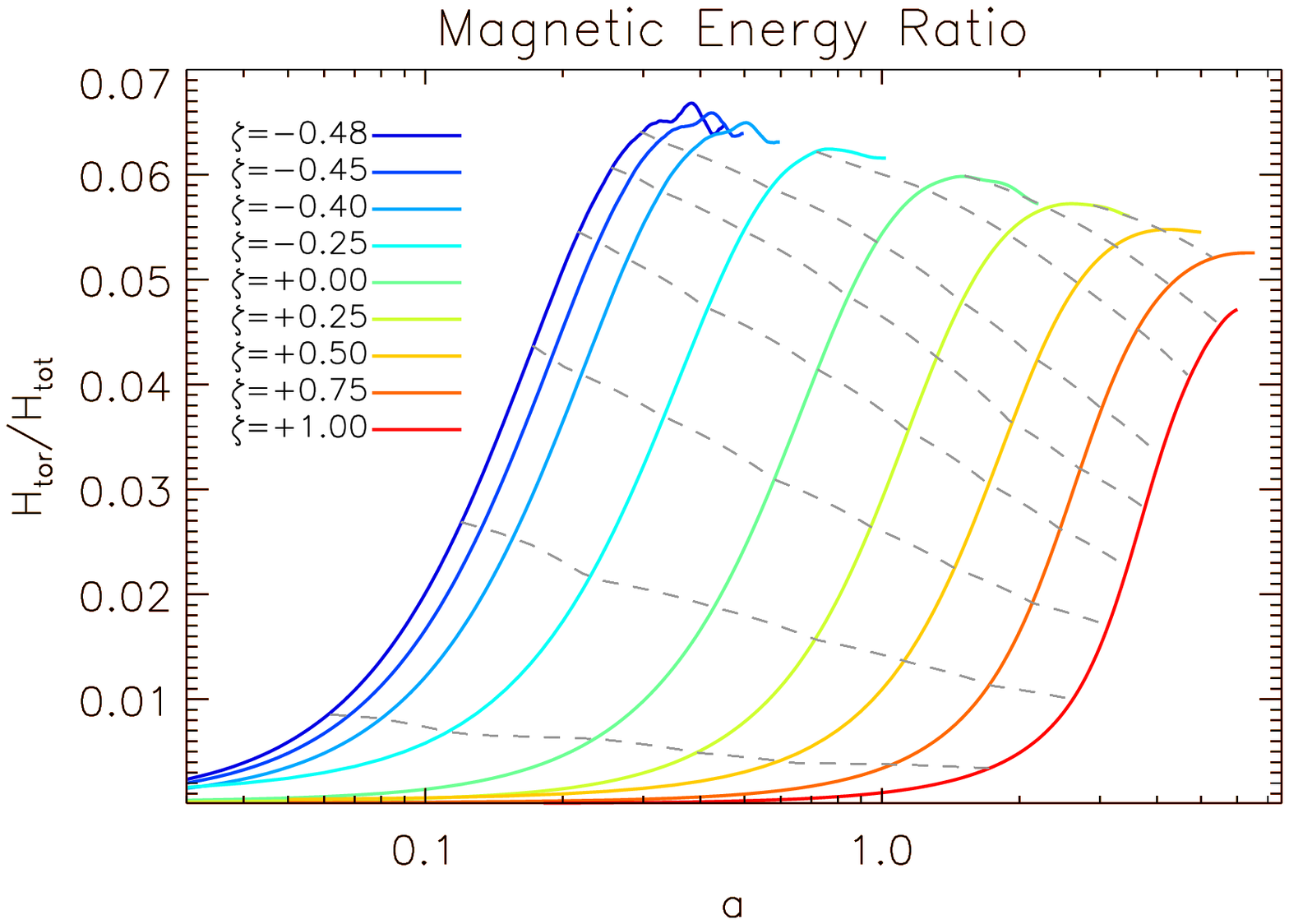}
\caption{Value of the ratio $\mathcal{H}_{\rm tor}/\mathcal{H}$ for TT
  sequences characterized by different values for 
  $\zeta$ as a function of $a$. The dashed lines correspond to
  configurations where the ratio between the maximum strength  of the toroidal magnetic field $B_{\rm
    tor}^{\rm max}$, and the maximum strength of the poloidal
  component $B_{\rm pol}^{\rm max}$ is constant. From  bottom to top
  $B_{\rm  tor}^{\rm max}/B_{\rm pol}^{\rm max} = 0.1,0.2,0.3,0.4,0.5,0.6,0.8,1.0,1.25$.
}
\label{fig:ttrat}
\end{figure}

\begin{figure*}
	\centering
	\includegraphics[width=.33\textwidth, bb=0 0 452 440, clip]{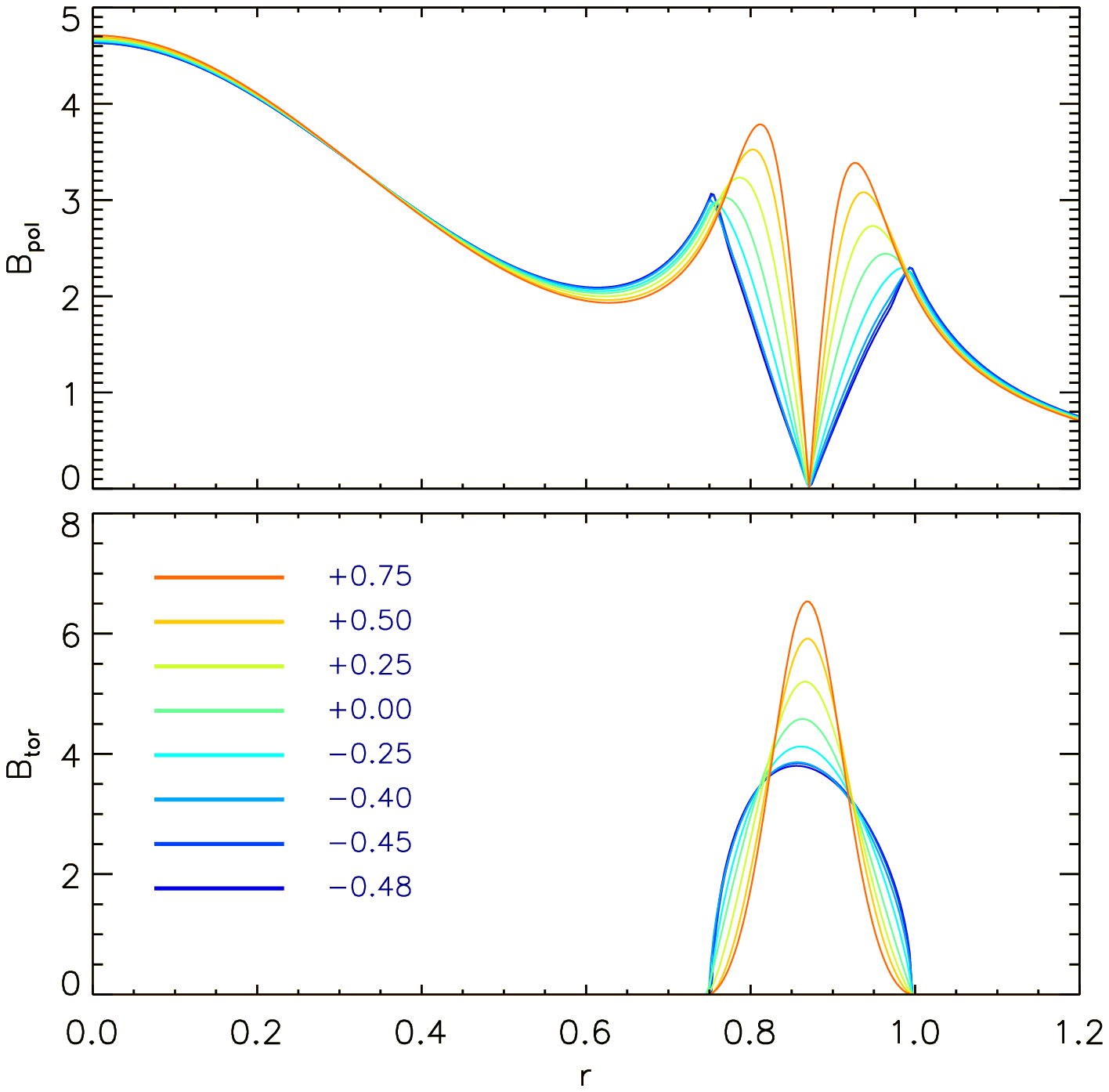}
	\includegraphics[width=.33\textwidth, bb=0 0 452 440, clip]{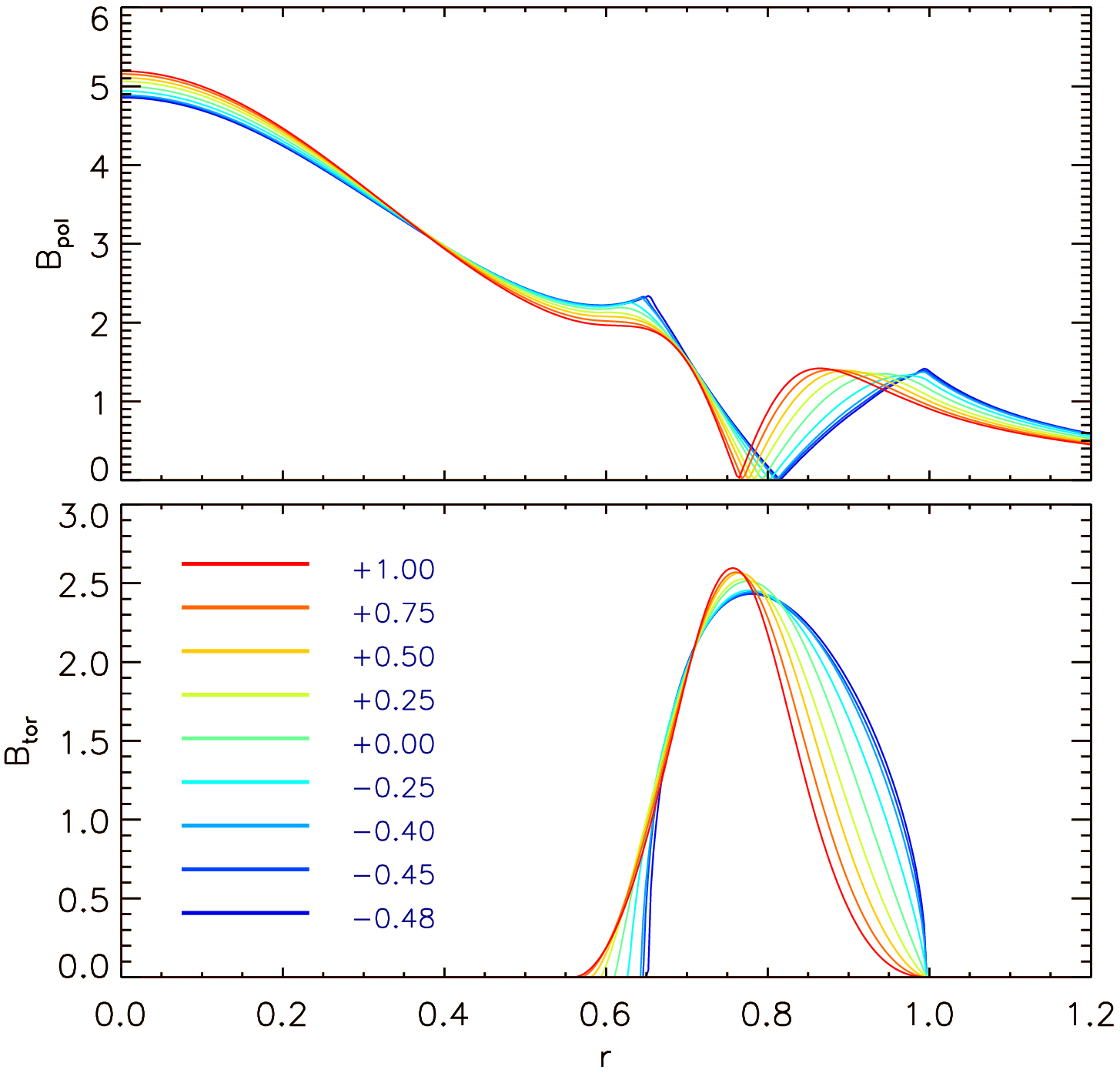}
	\includegraphics[width=.33\textwidth, bb=0 0 452 440,  clip]{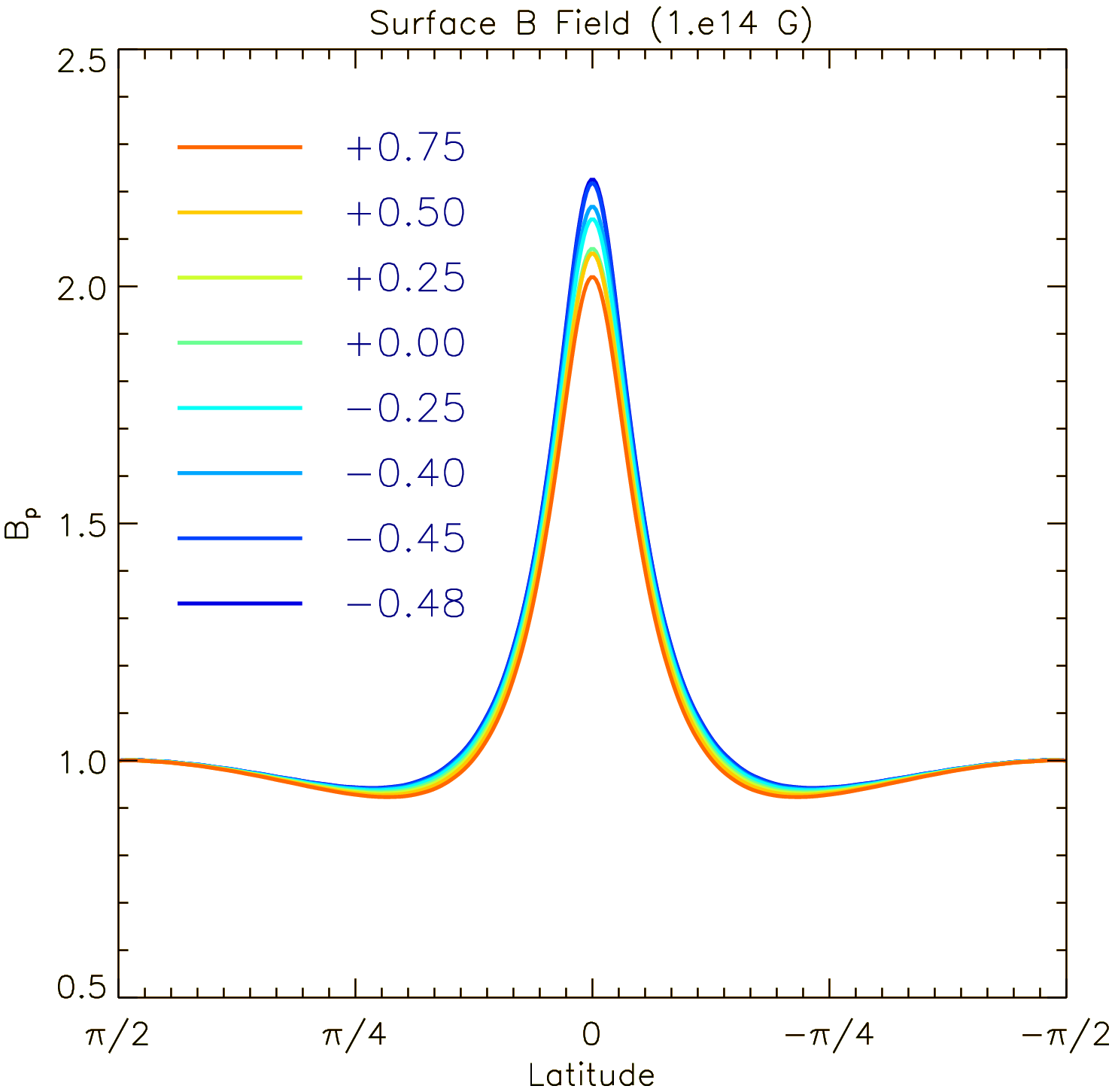}\\
\caption{Left panel: strength of the poloidal magnetic field (top) and
  toroidal magnetic field (bottom) inside the star, on the
  equatorial plane, as a function of $r$, normalized to $R_{\rm NS}$, for models corresponding to
  the maximum of $\mathcal{H}_{\rm tor}/\mathcal{H}$, for various
  values of $\zeta$. 
  Middle panel:strength of the poloidal magnetic field (top) and
  toroidal magnetic field (bottom) inside the star, on the
  equatorial plane, as a function of $r$, for models corresponding to
  $B_{\rm  tor}^{\rm max}/B_{\rm pol}^{\rm max} =0.5$. Right panel:
  strength of the magnetic field at the surface  for models corresponding to
  the maximum of $\mathcal{H}_{\rm tor}/\mathcal{H}$. In all cases
  the strength is normalized to the surface value at the pole.  
}
\label{fig:ttsur}
\end{figure*}

In Fig.~\ref{fig:ttrat} we show how the ratio of magnetic energy associated to
the toroidal field  $\mathcal{H}_{\rm tor}$ over the total magnetic
energy $\mathcal{H}$ changes with the
parameter $a$ and $\zeta$. The maximum value of this ratio is always of the order of 0.06,
slightly higher for smaller values of $\zeta$. In all cases we
verified that at high values of $a$ the volume of the region
containing the toroidal magnetic field is strongly reduced. For
$\zeta=1$ we could not find equilibrium models (solution of the GS
equation)  all the way to the maximum (the
algorithm failed to converge). Given that, for Eq.~\eqref{eq:fbern1},
both the energy of toroidal magnetic
field and the associated current scale with $\mathcal{I}$, one cannot
increase one without increasing the other. The systems seem always to
self-regulate, with a maximum allowed current, implying a maximum
allowed toroidal magnetic energy. The value of $\zeta$ affects the local
value and distribution of the magnetic field, but does not play a relevant
role for integrated quantities like currents and magnetic
energy. Indeed by looking at Fig.~\ref{fig:ttrat}, and Fig.~\ref{fig:ttsur}, it is
evident that for $\zeta<0$ it is not possible to have configurations
where the maximum strength of the toroidal field exceeds the one of the
poloidal field. For smaller $\zeta$ the same toroidal magnetic field
energy, corresponds in general to weaker toroidal magnetic fields. For
$\zeta>0$ instead we could reach configurations with a toroidal field
stronger than the poloidal one. Interestingly the volume of the torus,
for configurations where the ratio  $\mathcal{H}_{\rm
  tor}/\mathcal{H}$ is maximal,
does not depend on $\zeta$. 

One can also look at the magnetic field
distribution on the surface of the star. Given our previous results
for purely poloidal configurations with nonlinear current terms, we expect
strong deviations from the standard dipole, where the strength of the magnetic field
at the pole is twice the one at the equator. In Fig.~\ref{fig:ttsur} we show
the total strength of the magnetic field at the surface (where the field
is purely poloidal), for configurations where the ratio
$\mathcal{H}_{\rm tor}/\mathcal{H}$ is maximal. The presence of a
current torus, just underneath the surface, is evident in the peak of the
field strength at the equator. The peak is even narrower than what was
found for purely poloidal cases with $\xi=10$, and the strength of
the equatorial field can be more than twice the polar one. Again, there
is little difference among cases with different $\zeta$. Higher values
of 
$\zeta$ correspond to currents that are more concentrated around the
neutral line, located at $\sim 0.85 R_{\rm NS}$, and as such buried deeper
within the star. Indeed the strength of the magnetic field at the
equator with respect to the value at the pole,  is higher for smaller $\zeta$.

Recently CR13 have presented results where the
ratio $\mathcal{H}_{\rm tor}/\mathcal{H}$ is $> 0.5$ and can reach
value close to unity. However, in all of our models we get values
$\mathcal{H}_{\rm tor}/\mathcal{H}$ always less than 0.1.

A precise comparison with CR13, is non trivial. For example, using the definition of current in their
Eq.3, does not lead to converged solutions in the purely poloidal
case (confirmed by Ciolfi, private communication). This because their formulation of Eq.3, with a non-linear term
which introduces a subtractive currents with respect to the linear
one, can
lead to current inversions inside the NS. As we pointed out, our
algorithm fails (diverges) every time we attempt to model systems with
current inversions, and this might be related with uniqueness issue of
the elliptical Grad-Shafranov equation. If this is indeed an issue
with uniqueness then different numerical approaches might be more or
less stable, and the robustness of the solution becomes questionable.

 Note that CR13 impose that the field at the surface is a pure dipole,
 setting all other multipoles to zero. This might probably filter out
 and suppress the formation of localized currents at the edge of the NS
 and any effect associated to small scale structures, like the increase
 of the value of $A^\phi_{sur}$. As we show in this paper, the structure
 of the magnetic field at the surface, can dramatically differ from a
 pure dipole, depending on the current distribution. Even using the
 functional form by CR13, in the range where our code converges, we
 found that at the surface of the NS the magnetic field is far from a
 pure dipole.

 Imposing a purely dipolar field outside the stellar surface may have
 been determinant in the results of CR13, but because we are not able to
 impose such a boundary condition, further independent verification is
 needed to resolve this issue.

\subsection{Twisted Ring Configurations}
\label{sec:tr}

\begin{figure}
	\centering
	\includegraphics[width=.50\textwidth]{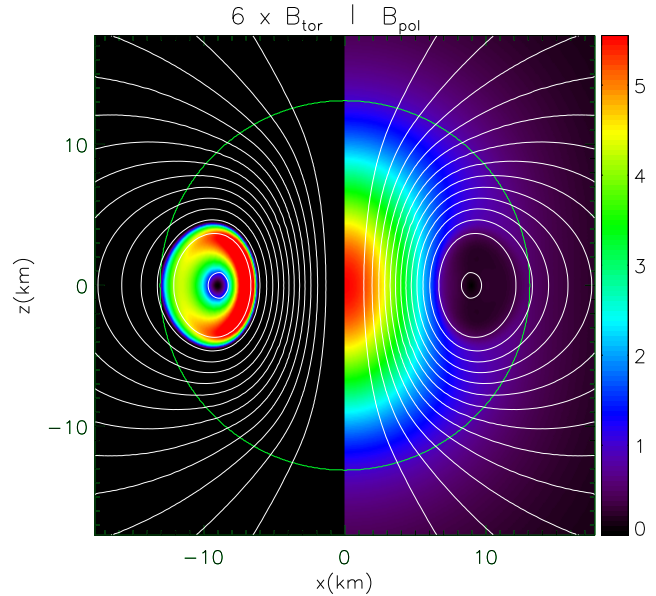}
\caption{Magnetic field for a TR configuration with
  $\zeta=0$ and $a=12.6$ (corresponding to a ratio
  $B_{\rm  tor}^{\rm max}/B_{\rm pol}^{\rm max} =0.15$ close to the
  maximum). Strength of the toroidal magnetic field (left) multiplied
  times a factor 6 for convenience, and poloidal
  magnetic field (right) normalized to the surface value at the
  pole. White contours represent 
			magnetic field surfaces (isocontours of
                        $A_\phi$).  The thick green line is the stellar
                        surface. Axes refer
                        to distances in a cartesian frame centered on
                        the origin and with the $z$-axis corresponding
                        to the symmetry axis.
}
\label{fig:tr}
\end{figure}

In the previous section we have shown that in the case of TT geometry it is
not possible to reach toroidally dominated configurations. This result
is also independent on the particular shape of the current
distribution $\mathcal{I}$. The system always self-regulates. As was
pointed out by CR13 this is due to the one to
one correspondence between integrated quantities, like the net current
and magnetic field energy. Motivated by
this, we can look for different forms for the equation $\mathcal{I}$
that allow a larger toroidal field, with a smaller net integrated
current. The current given by Eq.~\eqref{eq:fbern1} has always the same
sign, and as shown, acts as an additive term. 
On the other hand, the current associated to Eq.~\eqref{eq:fbern2}
changes its sign within the toroidal region where it is defined. The
field in this case has a geometry reminiscent of a \textit{Twisted
  Ring} TR: its strength vanishes on the neutral line, where also
the poloidal field goes to zero, and reaches a maximum in a shell
around it. This can be clearly seen in Fig.~\ref{fig:tr}. The net
integrated currents in this case, is much less than in the case of
Eq.~\eqref{eq:fbern1}, and it is globally subtractive. 

\begin{figure}
	\centering
	\includegraphics[width=.50\textwidth]{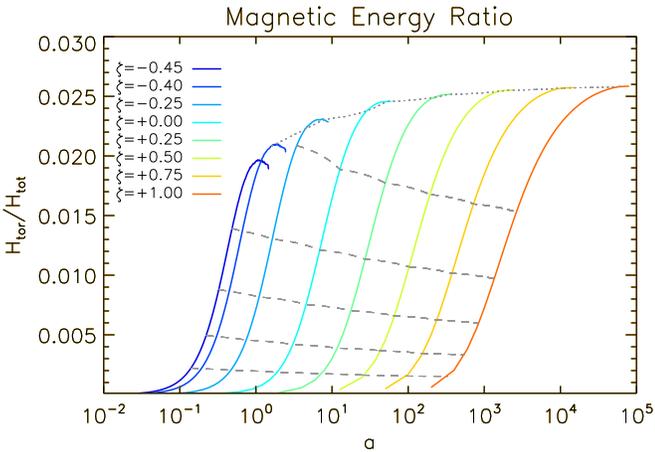}
\caption{Value of the ratio $\mathcal{H}_{\rm tor}/\mathcal{H}$ for TR
  sequences characterized by different values for 
  $\zeta$ as a function of $a$. The dashed lines correspond to
  configurations where the ratio between the maximum strength  of the toroidal magnetic field $B_{\rm
    tor}^{\rm max}$, and the maximum strength of the poloidal
  component $B_{\rm pol}^{\rm max}$ is constant. From bottom to top
  $B_{\rm  tor}^{\rm max}/B_{\rm pol}^{\rm max} =
  0.05,0.075,0.10,0.125,0.150$. The dotted line corresponds to
  configurations where $B_{\rm  tor}^{\rm max}/B_{\rm pol}^{\rm max} =
  0.14$, indicating that the ratio of the magnetic field component is
  not monotonic.
}
\label{fig:trrat}
\end{figure}

In Fig.~\ref{fig:trrat} we show how the ratio of magnetic energy associated to
the toroidal field  $\mathcal{H}_{\rm tor}$ over the total magnetic
energy $\mathcal{H}$ changes with the parameter $a$ and $\zeta$. Again
we find that it is not possible to build models that are toroidally
dominated. The maximum value of the ratio $\mathcal{H}_{\rm
  tor}/\mathcal{H}$ never exceeds 0.03 for all the values of $\zeta$
that we have investigated. The reason now is exactly the opposite of
the one for TT configurations. The current of TR geometry, as
anticipated, is subtractive. It acts like the nonlinear terms in the
purely poloidal configurations with $\xi <0$. Its effect is to remove
current from the interior of the star. This means that in the region
where $\mathcal{I} \neq 0$, the vector potential $A_\phi$ becomes
shallower: the quantity $[A_{\phi}^{\rm max}-A_{\phi}^{\rm sur}]$
diminishes. However, the strength of the toroidal magnetic field itself 
scales as $[A_{\phi}^{\rm max}-A_{\phi}^{\rm sur}]$. The nonlinearity
of the problem manifests itself again as a self-regulating
mechanism. Increasing $a$, in principle, implies a higher subtractive
current, but this reduces the value of $[A_{\phi}^{\rm
  max}-A_{\phi}^{\rm sur}]$, and the net result is that subtractive
current saturates, and the same holds for the toroidal magnetic
field. This saturation is reached at small values of $\mathcal{H}_{\rm
  tor}/\mathcal{H}$. Indeed, in Fig.~\ref{fig:trrat}, a clear maximum
is only visible for $\zeta <0$, while for $\zeta \ge 0$ the curves
seem to saturate to an asymptotic value. Again we find that the value
of $\zeta$ leads to small variations, with higher values of $\zeta$
leading to configurations with  slightly higher value of $\mathcal{H}_{\rm
  tor}/\mathcal{H}$. 

In all the parameter space we have investigated the strength of the
toroidal magnetic field never exceeds the one of the poloidal
component. At most, the toroidal magnetic field reaches values that are
$\sim 0.15$ times the maximum value of the poloidal field. This is in
sharp contrast with what was found for TT cases. Moreover, while in
the TT cases the maximum strength of the toroidal field  $B_{\rm
    tor}^{\rm max}$ was found to be a monotonically increasing function
  of the parameter $a$, along sequences at fixed $\zeta$, now $B_{\rm
    tor}^{\rm max}$ reaches a maximum $\sim 0.15 B_{\rm
    pol}^{\rm max}$, and then slowly diminishes, as can be seen from
  Fig.~\ref{fig:trrat}. This is again a manifestation of the effect of
  subtractive currents. Interestingly, the region occupied by the
  toroidal magnetic field does not shrink as $a$ increases. The
  saturation of the toroidal magnetic energy is not due to a reduction
  of the volume filled by the toroidal field, but to a depletion of
  the currents.

\begin{figure*}
	\centering
	\includegraphics[width=.33\textwidth, bb=0 0 452 440, clip]{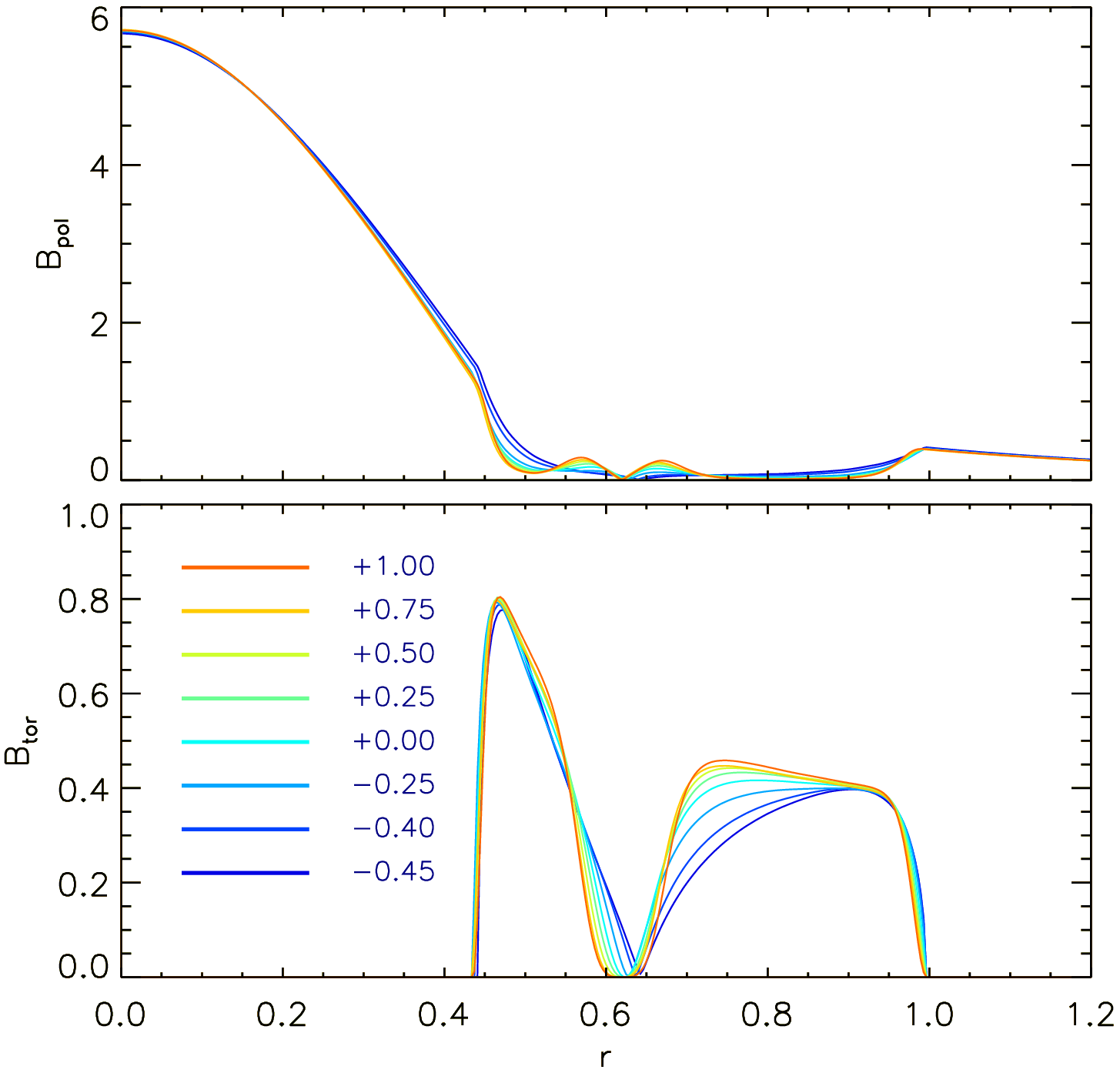}
	\includegraphics[width=.33\textwidth, bb=0 0 452 440, clip]{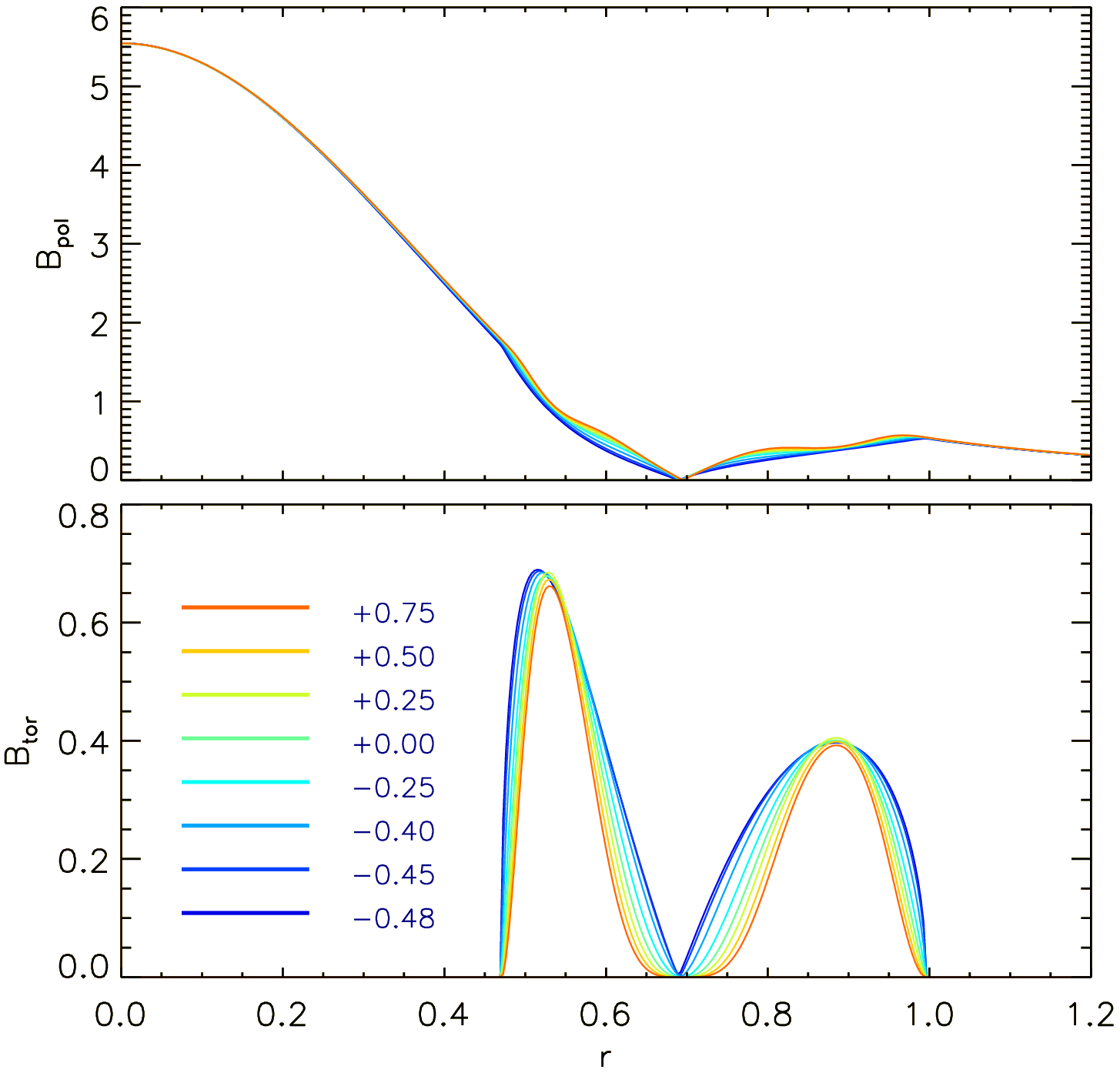}
	\includegraphics[width=.33\textwidth, bb=0 0 452 440,  clip]{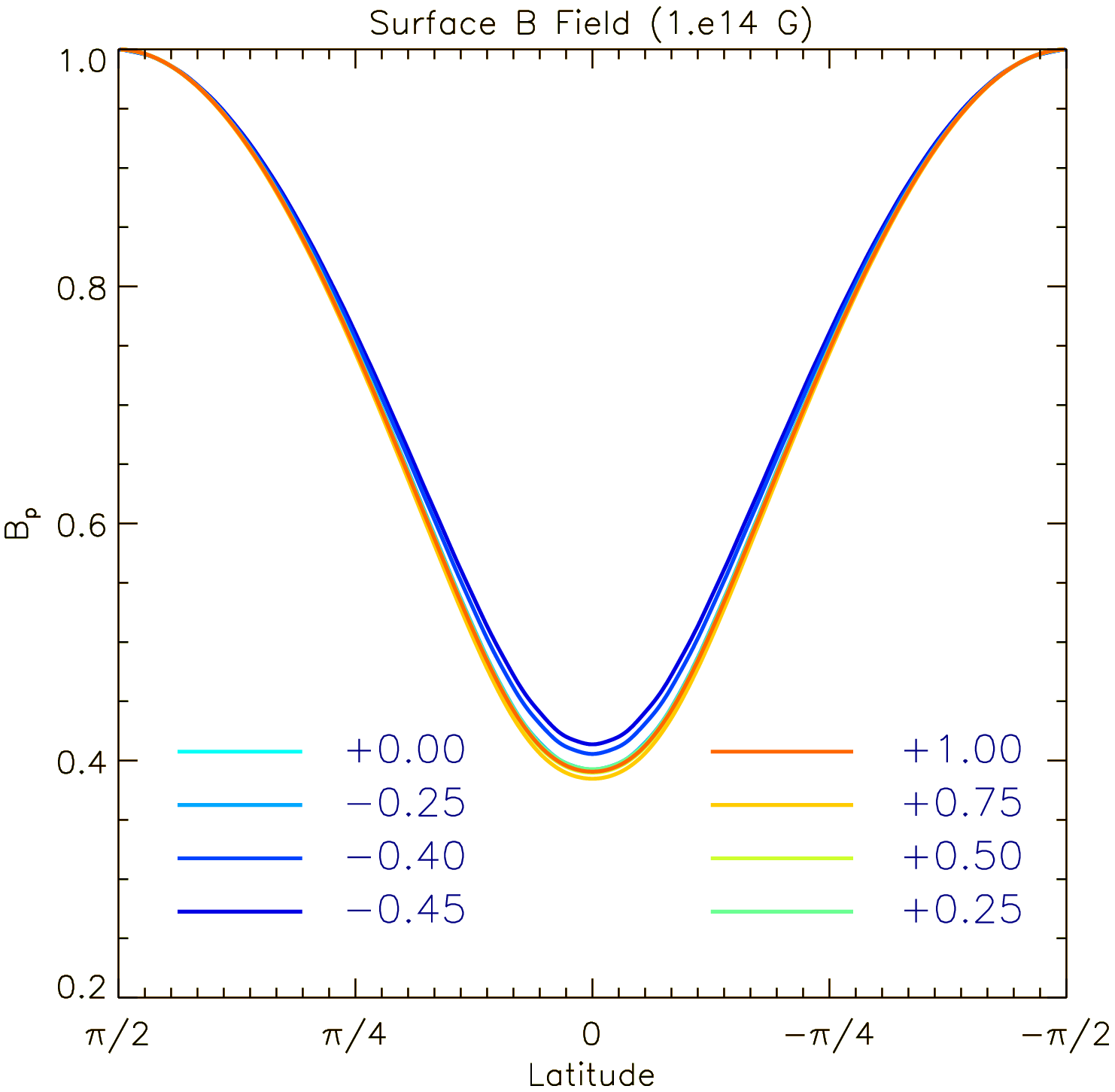}\\
\caption{Left panel: strength of the poloidal magnetic field (top) and
  toroidal magnetic field (bottom) inside the star, on the
  equatorial plane, as a function of $r$, normalized to $R_{\rm NS}$,for models corresponding to
  the maximum of $\mathcal{H}_{\rm tor}/\mathcal{H}$, for various
  values of $\zeta$. 
  Middle panel:strength of the poloidal magnetic field (top) and
  toroidal magnetic field (bottom) inside the star, on the
  equatorial plane, as a function of $r$, for models corresponding to
  $B_{\rm  tor}^{\rm max}/B_{\rm pol}^{\rm max} =0.5$. Right panel:
  strength of the magnetic field at the surface  for models corresponding to
  the maximum of $\mathcal{H}_{\rm tor}/\mathcal{H}$. In all cases
  the strength is normalized to the surface value at the pole.  
}
\label{fig:trsur}
\end{figure*}

As was done for the TT cases, we can also look at the distribution of
magnetic field inside the star. In Fig.~\ref{fig:trsur}, we show the
strength of the poloidal and toroidal components of the magnetic field
along an equatorial cut. The effect of subtractive currents is
evident in the suppression of the poloidal field in the TR region
that extends from about half the star radius to its outer edge. It is also
evident that the value of $\zeta$ plays only a minor role, and that
differences are stronger at saturation than for intermediate
values. Interestingly, there are very marginal effects concerning the
strength of the magnetic field at the surface, which is essentially the
same as the standard dipole. Again this can be partially understood
recalling the behaviour of purely poloidal configurations with $\xi
<0$. In those cases, substantial deviations from the dipolar case
were achieved only in the limit $\xi \rightarrow 1$, when a large part
of the star was unmagnetized. Here the size of the unmagnetized ring
region remains more or less constant, and it does not affect the
structure of the field at the surface. The global effect of the
subtractive currents is small, and this reflects in the trend of
the magnetic dipole moment, which diminishes only slightly  by about 30-40\%.

\subsection{Dependence on the stellar model}
\label{sec:mass}

In the previous sections we have investigated in detail the role of
two families of currents $\mathcal{I}$, that can be considered quite
representative of a large class of current configurations. Our results
show that in neither case we could obtain magnetic field distributions
where the energetics was dominated by the toroidal component. 

In this section we try to investigate the importance of the underlying
stellar model. In general, previous studies have mainly focused on the
distribution of currents, assuming a reference model for the NS: either
a $1.4 M_\odot$
\citep{Ciolfi_ferrari+09a,Ciolfi_Ferrari+10a,Ciolfi_Rezzolla13a,Lander_Jones09a}
  or a $1.5 M_\odot$
  \citep{Pili_Bucciantini+14a} NS. Only
  \citet{Glampedakis_Andersson+12a} have partly investigated how
  the stellar structure might affect the energetics properties of the
  magnetic field. In particular they focused on the role of
  stable stratification, and showed that this might change the maximum
  amount of magnetic energy associated to the toroidal magnetic
  field, in standard TT configurations. 

\begin{figure}
	\centering
	\includegraphics[width=.50\textwidth]{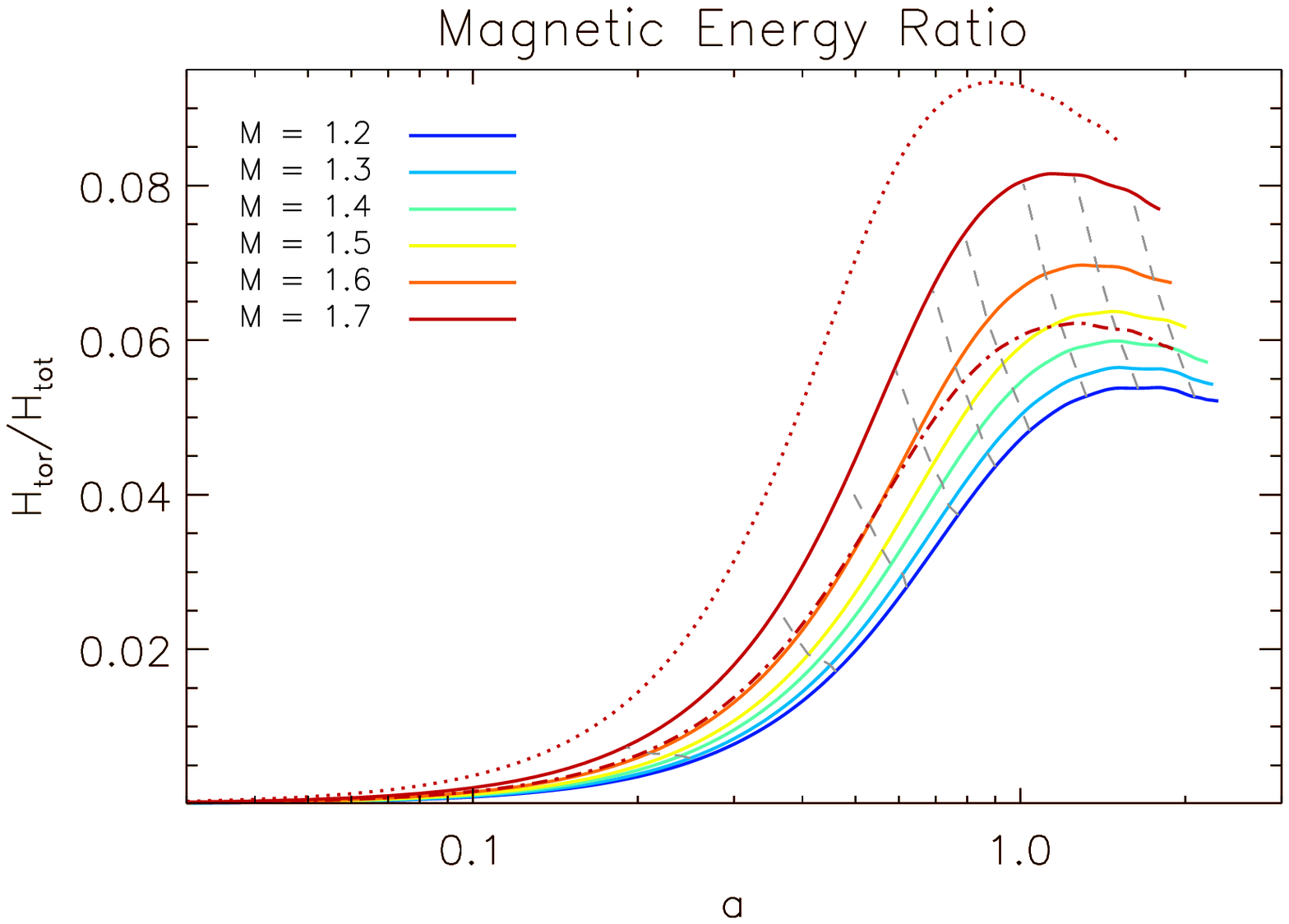}
\caption{Value of the ratio $\mathcal{H}_{\rm tor}/\mathcal{H}$ for TT
  sequences with   $\zeta=0$, characterized by different values for the gravitational mass
as a function of $a$. The dashed grey lines corresponds to
  configurations where the ratio between the maximum strength  of the toroidal magnetic field $B_{\rm
    tor}^{\rm max}$, and the maximum strength of the poloidal
  component $B_{\rm pol}^{\rm max}$ is constant. From bottom to top
  $B_{\rm  tor}^{\rm max}/B_{\rm pol}^{\rm max} =
  0.1,0.2,0.3,0.4,0.5,0.6,0.8,1.0,1.25$. The dotted red line
  corresponds to a
  configuration with $M=2.0M_\odot$, and the same central rest mass density as
  the $1.7 M_\odot$ case. The dot-dashed red line
  corresponds to a
  configuration with $M=1.7M_\odot$, but a lower central rest mass density
  with respect to 
  the $K_{\rm a}=110$ case.
}
\label{fig:massrat}
\end{figure}

In Fig.~\ref{fig:massrat} we show how the ratio $\mathcal{H}_{\rm
  tor}/\mathcal{H}$  changes as a function of $a$ for standard TT
models with $\zeta=0$, but for NSs with different masses. For $K_{\rm
  a}=110$ the maximum mass for a NS is found to be $\sim 1.7
M_\odot$. It is clear that models with a higher mass have a higher
value of the ratio $\mathcal{H}_{\rm
  tor}/\mathcal{H}$ , for the same value of $a$. Interestingly, the
maximum value reached by $\mathcal{H}_{\rm
  tor}/\mathcal{H}$ for a  $1.7 M_\odot$ NS, is about 0.08, compared to
0.06 for a $\sim 1.4 M_\odot$ NS. This is a substantial relative
increase, even if the magnetic energy is still dominated by the
poloidal component. Moreover this increasing trend is stronger at
higher masses. 

We also investigated how much of this trend is related just to the total
stellar mass (i.e. the compactness of the system) and how much depends
on the value of rest mass density in the core
of the NS. Indeed it was previously found the NSs with higher masses can
harbor in principle stronger magnetic fields
\citep{Pili_Bucciantini+14a}. On the other hand, the current
associated with $\mathcal{M}$, responsible for the structure of the
poloidal field, scales as the rest mass density. For models built by keeping constant $K_{\rm
  a}=110$, a higher mass implies a higher central rest mass density, so that 
it is hard to disentangle them. In Fig.~\ref{fig:massrat} we show also
two models with different EoS: one that has the same central rest mass
density as the  $1.7 M_\odot$
NS, but different values of the adiabatic constant $K_{\rm
  a}$, such that is total gravitational mass is   $2.0 M_\odot$; the
other has the same mass of $1.7 M_\odot$, but a lower central rest mass
density (about one third). It is evident that models with a smaller total mass, given
the same central rest mass density, correspond to lower maximum value for $\mathcal{H}_{\rm
  tor}/\mathcal{H}$. On the other hand, given the same central rest
mass density,
the ratio $\mathcal{H}_{\rm
  tor}/\mathcal{H}$ clearly increases with total mass. It appears that
the rest mass density stratification (how much concentrated is the
rest mass density
distribution in the core and how much shallow is it in the outer
layers), regulates the relative importance of $\mathcal{I}$ and
$\mathcal{M}$, and the net outcome in terms of energetics of the
toroidal and poloidal components.

\subsection{Mixed nonlinear currents}
\label{sec:mix}

It was suggested by CR13 that a possible reason
why TT configurations, computed using $\xi=0$ in $\mathcal{M}$, could
not achieve the toroidally dominated regime, was due to the fact that
the contribution to the azimuthal current from $\mathcal{I}$ soon
dominates. As a consequence, the resulting poloidal configuration enters the
nonlinear regime in which the size of the torus region, where the toroidal
field is confined, shrinks.
They show that, by introducing a current term in
$\mathcal{M}$ to compensate for $\mathcal{I}$, it was possible to
avoid this behaviour. However they also stressed the fact that a very
peculiar form for  $\mathcal{M}$ was needed to achieve significative
results.

\begin{figure*}
	\centering
	\includegraphics[width=.33\textwidth, bb=0 0 492 440, clip]{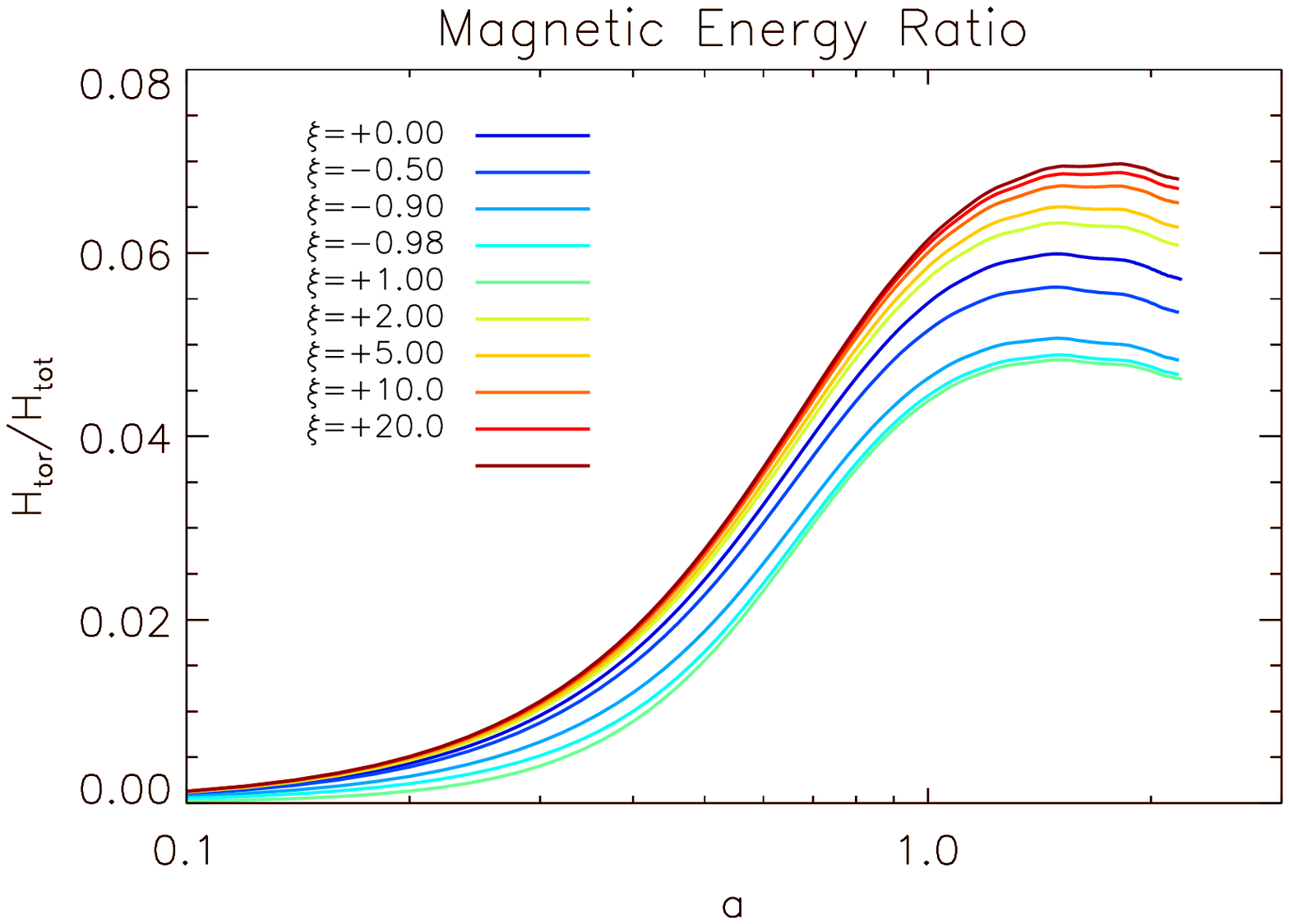}
	\includegraphics[width=.33\textwidth, bb=0 0 492 440, clip]{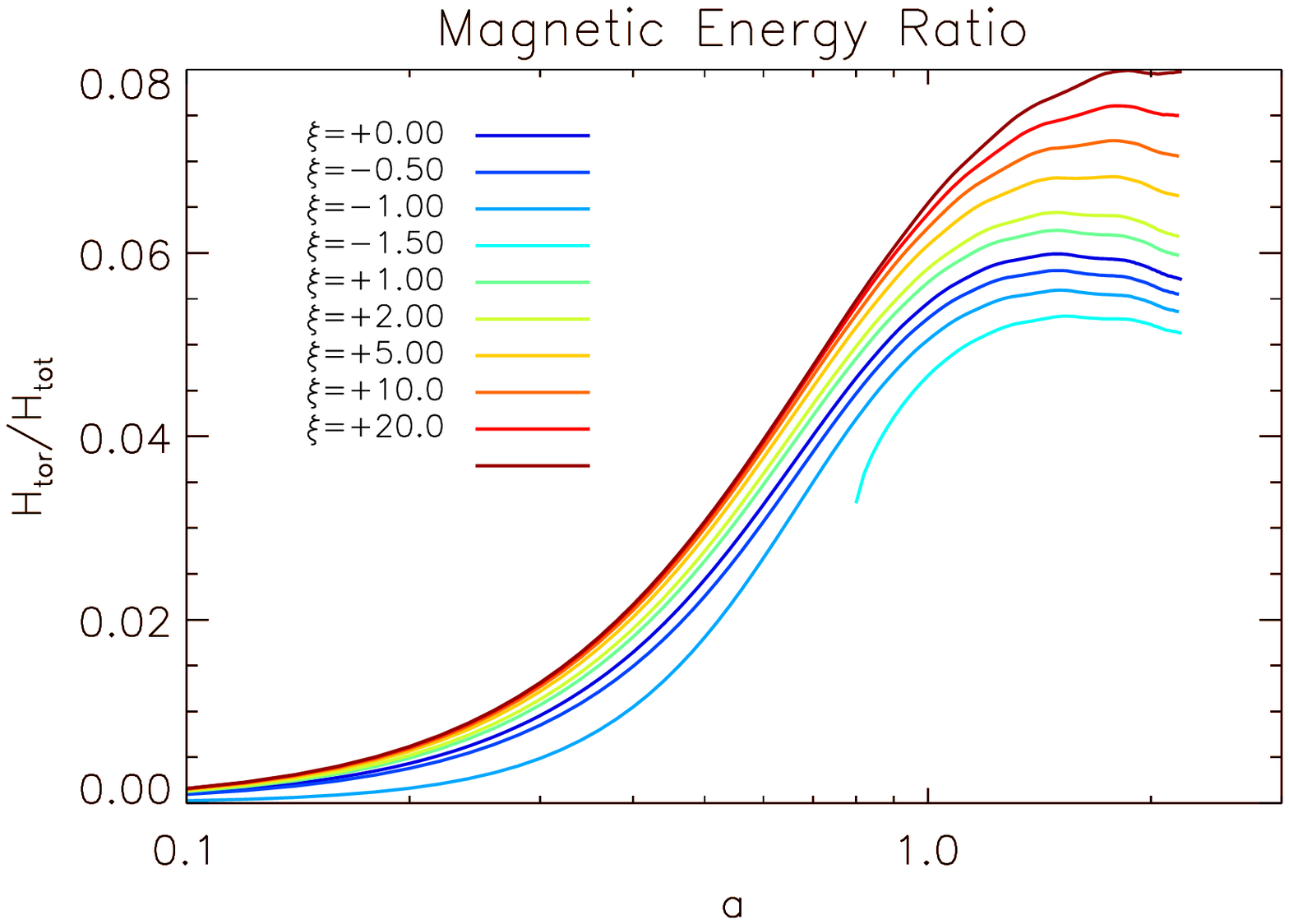}
	\includegraphics[width=.33\textwidth, bb=0 0 492 440,  clip]{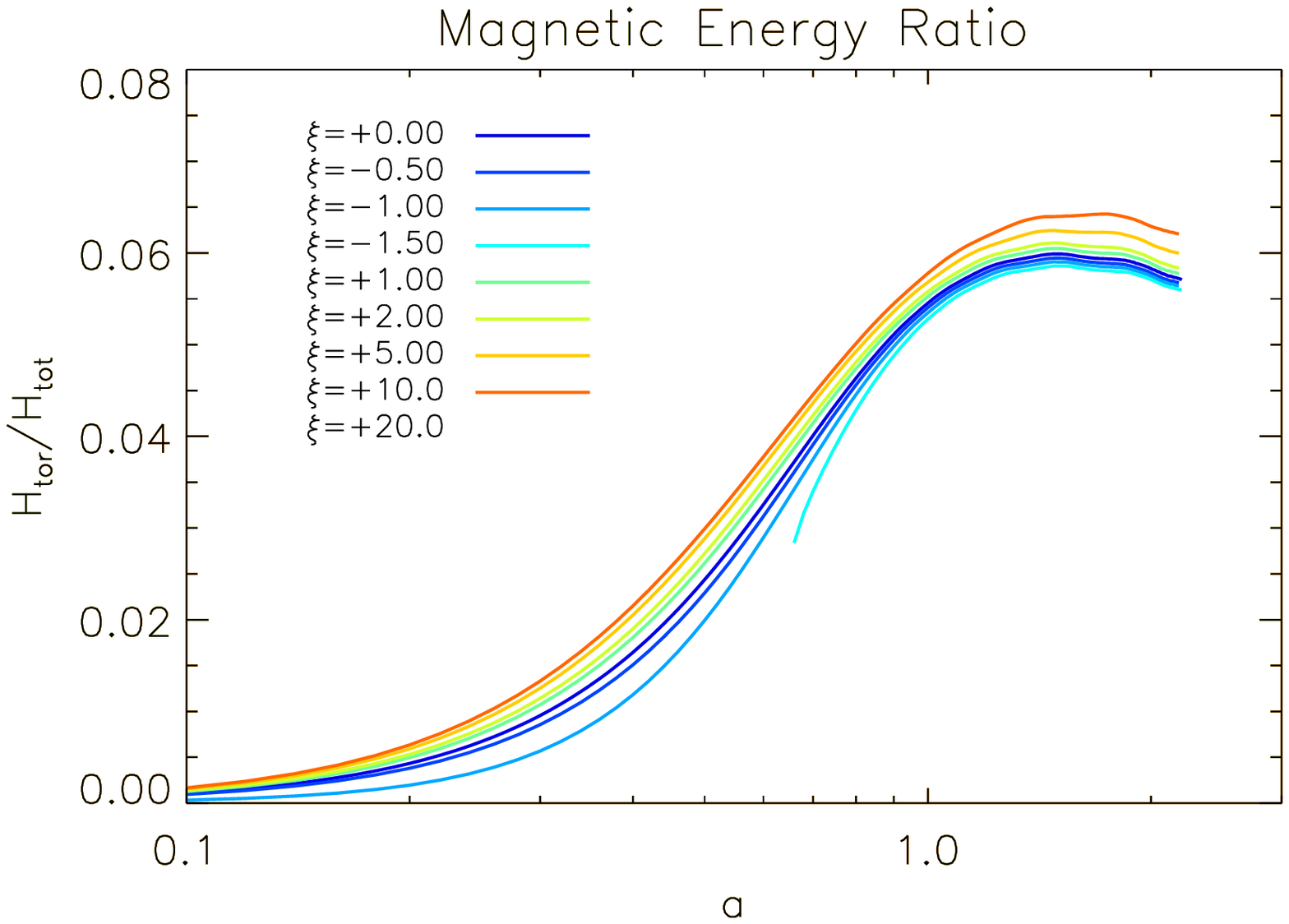}\\
\caption{Values of the ratio $\mathcal{H}_{\rm tor}/\mathcal{H}$ for
  TT configurations with $\zeta=0$, in the presence of nonlinear terms in the definition of
  $\mathcal{M}$.
 Left panel: cases with $\nu=1$. Middle panel: cases with
 $\nu=4$. Right panel: cases with $\nu=10$.
}
\label{fig:nl}
\end{figure*}

Here we investigate what happens to TT models, using for $\mathcal{I}$ the
form of Eq.~\eqref{eq:fbern1}, if we retains nonlinear
terms in the definition of $\mathcal{M}$, and what happens in cases where
$\xi \neq 0$. In Fig.~\ref{fig:nl} we show how $\mathcal{H}_{\rm
  tor}/\mathcal{H}$ changes for TT configurations with $\zeta=0$ for
various values of the parameter $\xi$ and for selected values of
$\nu=1,4,10$. Naively, based on the idea that compensating currents
are needed to achieve toroidally dominated configurations, 
one would expect that higher values of $\mathcal{H}_{\rm
  tor}/\mathcal{H}$ should be reached for $\xi <0$ (subtractive
currents). Fig.~\ref{fig:nl}  shows instead that the trend is the
opposite. In general, lower values of $\mathcal{H}_{\rm
  tor}/\mathcal{H}$ are found for $\xi<0$ and higher for $\xi>0$, even
if this is just a minor difference. The value of $\nu$ seems not to
play a major role. Interestingly the effect is maximal for
intermediate values of $\nu=4$, and marginal for $\nu=10$.

This counterintuitive trend is due to the fact that both the effects of
the current term $\mathcal{I}$ and the contribution of nonlinear
terms in $\mathcal{M}$, become important only in the fully nonlinear
regime. For values of $\xi \sim 0$ the effect of the nonlinear current
term in  $\mathcal{M}$ is negligible. For higher values of $\xi$ this
nonlinear term becomes more important. In the case $\xi<0$ they give
rise to a compensating current (the net dipole grows less) but, as
discussed, they also tend to suppress the vector potential and this effect
is stronger, leading to a overall decrease of the magnetic field. In
the case $\xi>0$, one would expect this additive current to lead to
an even more pronounced reduction in the torus volume, however, this is
not so. The net dipole increases but this additive currents enhance
the vector potential and the net result is a  higher $\mathcal{H}_{\rm
  tor}/\mathcal{H}$ (up to 30\% higher for $\nu=4$ and $\xi=20$).
The highly nontrivial behaviour of
the nonlinear regime is apparent. It is however possible that different forms 
for the compensating current might lead to different results.

Interestingly, again we are not able to construct equilibrium model
with current inversion. It is possible, for higher values of $\nu$,
to build models with $\xi <-1$, but only as long as the current in
the domain is always of the same sign. Indeed, cases with $\xi<-1$ are
allowed by the presence of a current due to $\mathcal{I}$, 
given by Eq.~\eqref{eq:fbern1}, that is always additive. 
There appears to be a threshold value for $a$ below which cases with
$\xi<-1$ are not realized. This is consistent with the argument about
local uniqueness we discuss in the purely poloidal case. Solutions with
subctractive currents can be built only as long as the nonlinear
current term is subdominant, and other currents enforce
stability. Given the presence of an extra current due to
$\mathcal{I}$, associated with the
toroidal magnetic field, now it is possible to build solutions with $\xi<-1$.

\begin{figure}
	\centering
	\includegraphics[width=.45\textwidth]{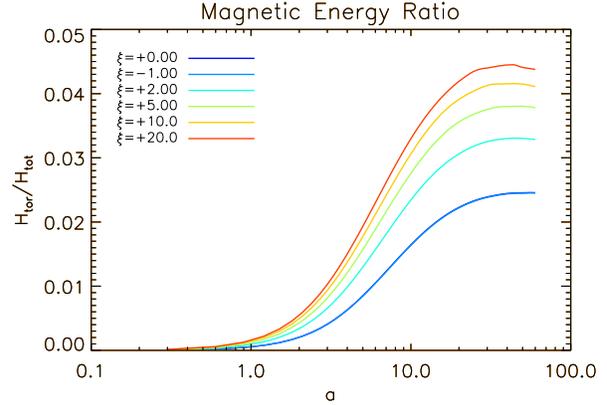}
\caption{Values of the ratio $\mathcal{H}_{\rm tor}/\mathcal{H}$ for
  TR configurations with $\zeta=0$, in the presence of nonlinear terms in the definition of
  $\mathcal{M}$, with $\nu=4$.
}
\label{fig:nl2}
\end{figure}

Similar results apply for the cases of TR configuration where
$\mathcal{I}$ is given by Eq.~\eqref{eq:fbern2}. In Fig.~\ref{fig:nl2}
we show these results. For values of
$\xi <0$ the ratio $\mathcal{H}_{\rm  tor}/\mathcal{H}$ is essentially
unchanged (it looks like the ratio is  marginally smaller).
 For positive values of $\xi$ we found a substantial
increase: $\mathcal{H}_{\rm  tor}/\mathcal{H}$ can be a factor 2
higher than in the simple TR case. In this case, the additive non
linear term in $\mathcal{M}$ compensates the subtractive current due
to $\mathcal{I}$, and stronger values for the magnetic field are
achieved. However, in the range of parameter investigated here, the ratio
$\mathcal{H}_{\rm  tor}/\mathcal{H}$ never exceeds 0.05. The energetics
is still dominated by the poloidal magnetic field.

Given the opposite behaviour of the currents associated with
$\mathcal{I}$, respectively from Eq.~\eqref{eq:fbern1}, and
Eq.~\eqref{eq:fbern2}, we also investigated configurations where the
current associated with $\mathcal{I}$, is given by a combination of TT
and TR configurations. Based on the results discussed above, we expect
that the additive term associated with the component of $\mathcal{I}$
from Eq.~\eqref{eq:fbern1}, should lead to results similar to what we
found for TR configurations with nonlinear terms in $\mathcal{M}$
with $\xi>0$. Indeed this is confirmed. In general we find that the ratio
$\mathcal{H}_{\rm  tor}/\mathcal{H}$ is smaller than for the TT
case, but larger than for TR case, even by a factor 2. It seems that
additive currents, at least for the functional form adopted here, tend to
dominate over subtractive ones.

\section{Conclusion}
\label{sec:conclusion}

In this work we investigated several equilibrium configurations
for magnetized NSs, carrying out a detailed study of the parameter
space. This allowed us to investigate general trends, and to
sample the role of various current distributions. Interestingly we
found that, almost insensitive of the chosen current distribution, the
ratio $\mathcal{H}_{\rm  tor}/\mathcal{H}$ never grows above 0.1.

 We tried to use the same prescription for the current structure inside
the star as the one used by CR13, but
we, not only could not reproduce their results, but we got the opposite
trend of a reduction in  $\mathcal{H}_{\rm
  tor}/\mathcal{H}$, consistent with all the other results we have
presented here. We pointed to a possible origin of this difference,
related perhaps to the choice in the boundary conditions
done by them,  but because we are not able to
 impose such a boundary condition, further independent verification is
 needed to resolve this issue.

The failure to get toroidally dominated configurations, that are
expected for stability in barotropic stars, might even point to the
possibility that barotropicity does not hold in NS, and the entire stability
problem is just related to entropy stratification
\citep{Reisenegger09a,Akgun_Reisenegger+13a}, and not to the current distribution: a stably
stratified NS can hold in place even a magnetic field out of MHD equilibrium.

On the other hand, the structure and strength of the magnetic field at
the surface, is strongly influenced by the location and distribution
of currents inside the star. We showed that magnetic field at the
equator can in principle be much higher or much smaller than the value
of the field at the pole. This means that the surface field can easily
be dominated by higher  multipoles than the dipole. It also implies
that local processes, at or near the surface, might differ substantially, in
their signatures, from the expectations of dipole dominated model, while
on the other hand, processes related to the large scale field, as
spin-down, will not. Interestingly, the result of the fully saturated
non-linear regime, in the presence of subtractive currents, looks
similar to what has recently been found in full time dependent MHD
simulation of core collapse and Proto-NS formation in Supernovae by
\citet{Obergaulinger_Janka+14a} (see the bottom panel of their Figure 14). The reason is due to the fact that turbulent eddies tend
to expel magnetic field \citep{Moffatt78a}, which concentrates toward the axis, and
becomes almost tangential at the proto-NS surface. Of course
turbulence introduces also small scales, which however are likely to
the first to be dissipated by any resistivity, leaving only the large
scale structure at later times.

We also showed that mass and central rest mass density can affect the energetic
properties of the magnetic field. In principle higher ratios of
$\mathcal{H}_{\rm  tor}/\mathcal{H}$ are reached for more massive and
denser NSs. This might suggest that magnetars are NSs with higher mass
than the average $1.4-1.5 M_\odot$. It also stresses the importance
and the role of the EoS, in determining possible electromagnetic
properties and signatures of the NS.

\section*{Acknowledgements}
This work has been supported by a EU FP7-CIG grant issued to the
NSMAG project (P.I. NB), and by the INFN TEONGRAV initiative 
(local P.I. LDZ).

\bibliography{my}{}
\bibliographystyle{mn2e}

\appendix
\section{Antisymmetric Solutions}.
\label{sec:anti}

As we already discussed in Sect.\ref{sec:poloidal}, the parity of the
magnetic field, with respect to the equator, depends on the parity of
the linear current term in the magnetization function
$\mathcal{M}$. All the solutions that we have shown previously are
symmetric (for $A_\phi$) with respect to the equator because this linear current term
is proportional to the rest mass density. This is a requirement built into the
integrability condition leading to the Bernoulli integral of Euler equation. It fixes the possible functional forms of
$\mathcal{M}$. If one is willing to relax the global integrability
condition, by allowing for example singular surface currents, it is
possible to obtain antisymmetric solutions. Due to the presence of a
surface current, there will be a jump in the parallel
component of the magnetic field at the surface. However, introducing
non linear current terms in $\mathcal{M}$, one can go to the fully non
linear saturated regime, where the contribution of the linear current
term becomes negligible, and make the residual jump in the magnetic
field at the surface arbitrarily small. The non linear current term
will preserve the parity of the surface current. We stress that, in
this case, equilibrium and integrability hold inside the star, except
at the surface itself.

\begin{figure}
	\centering
	\includegraphics[width=.45\textwidth]{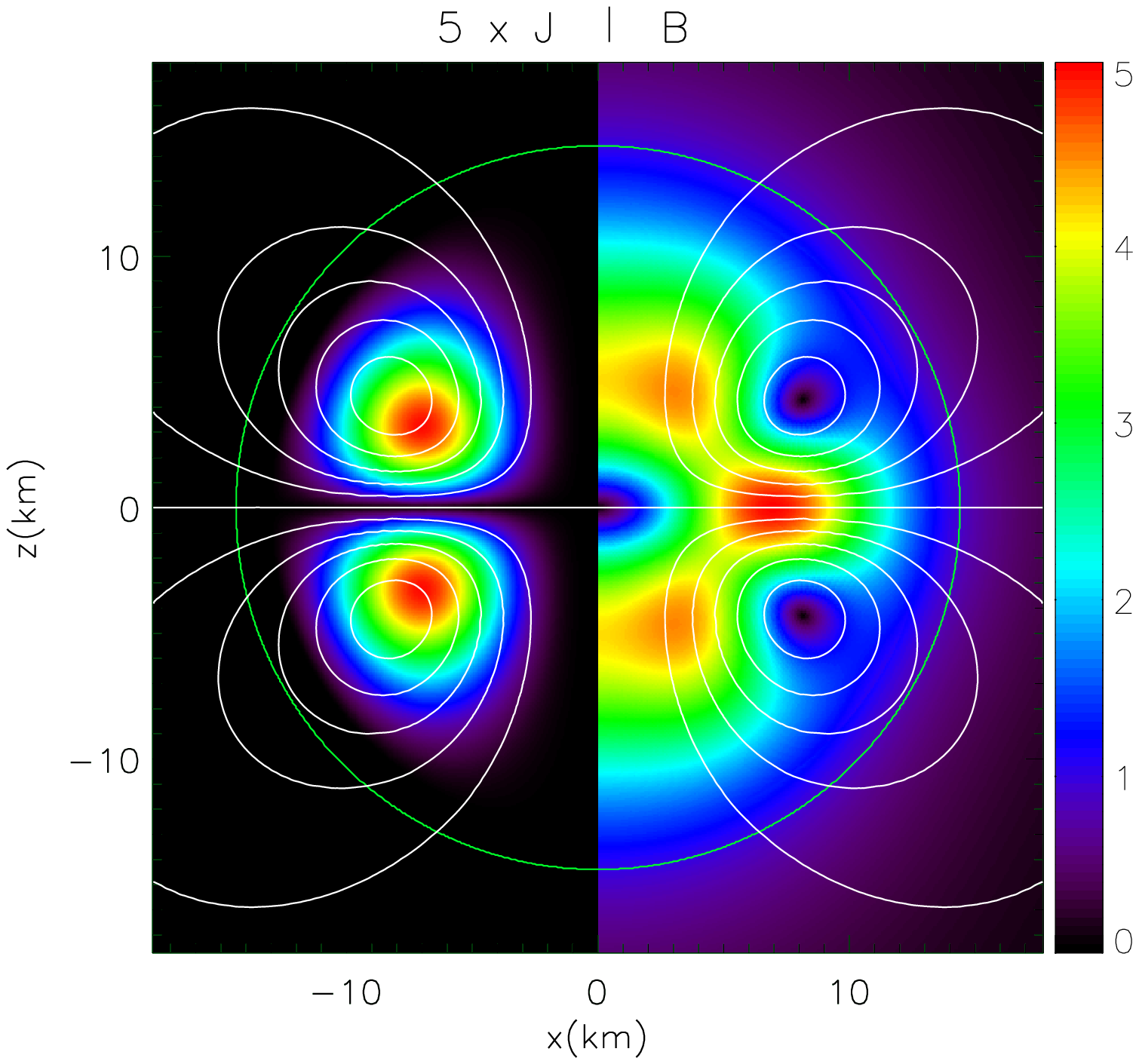}
\caption{Antisymmetric solution, with $\nu=1$. Left panel: azimuthal
  current density normalized to 0.2
  times its maximum. Right panel: magnetic field strength, normalized to the
  value at the pole. White contours represent 
			magnetic field surfaces (isocontours of
                        $A_\phi$).  The thick green line is the stellar
                        surface. Axes refer
                        to a cartesian frame centered on
                        the origin and with the $z$-axis corresponding
                        to the symmetry axis.
}
\label{fig:antisym}
\end{figure}

If we choose for $\mathcal{M}$ the following functional form:
\be
\mathcal{M}(A_\phi)=k_{\rm pol} A_\phi \left[\frac{\xi}{\nu+1}
\left(\frac{A_\phi}{A_{\phi}^{\rm max}}\right)^{\nu}\right],
\label{eq:mbernanti}
\ee
and add to the current $J^\phi$ [see Eq.~(\ref{eq:cur})], that enters the Grad-Shafranov
Eq.~(\ref{eq:gs}), a singular current term:
\be
k_{\rm pol}\cos{\theta}~\delta(r-R_{NS}),
\ee
by rising the value of $\xi$ one can find solutions that are
independent of the strength of the surface current. We show in
Fig.~\ref{fig:antisym} the result in the case $\nu=1$, $\xi=50$. The jump at the
surface is much smaller than the value of the magnetic field, and the solution
can be assumed to be smooth. The result is dominated by the
quadrupolar component. 

Note that the symmetry of the current term only
fixes the symmetry of the final solution. Every symmetric current will
lead to the same symmetric field, which depends only on $\nu$, while
every antisymmetric function will lead to the same antisymmetric
field, which again depends on $\nu$ alone. With this approach it is not
possible to produce for example octupolar models (where the dipole and
quadrupole components are absent). Even the use of an
octupolar surface current leads to dipolar configurations, in the
fully saturated nonlinear regime. In the presence of non linear
current term, multipoles are not eigenfunctions of the Grad
Shafranov, and mode mixing is introduced. For the values of $\nu$ that
we investigated, there is always a leading dipole component in the symmetric
case, and  a leading quadrupole component in the antisymmetric case,
even if the strength of higher order multipoles at the surface can be
relevant.

\section{Strong Field Regime}
\label{sec:strong}

Our formalism allows us to extend the solutions computed in the weak
field regime to the strong field regime to evaluate, for example, the related
deformation induced by the magnetic field. In the strong field regime,
however, the solution depends on the strength of the field. A detailed study of
the induced deformation in the case of a purely poloidal field with
$\xi=0$, and of TT configurations with $\zeta=0$, has already been
presented by \citet{Pili_Bucciantini+14a}. In that work there was also
an investigation of the role of non linear current terms in
$\mathcal{M}$, but only for $\nu=1$ and for small values of $\xi$ far
from the fully non linear saturated regime. The present results, about
TT configurations with various values of $\zeta$, show that
$\mathcal{H}_{\rm tor}/\mathcal{H}$ have similar trends to the
$\zeta=0$ case, and is always smaller than 0.1. We expect the
deformation to be similar to what was found in
\citet{Pili_Bucciantini+14a}.
On the other hand we have shown that, for  purely poloidal fields, the
non linear current term can substantially modify the field structure. 

\begin{figure}
	\centering
	\includegraphics[width=.45\textwidth]{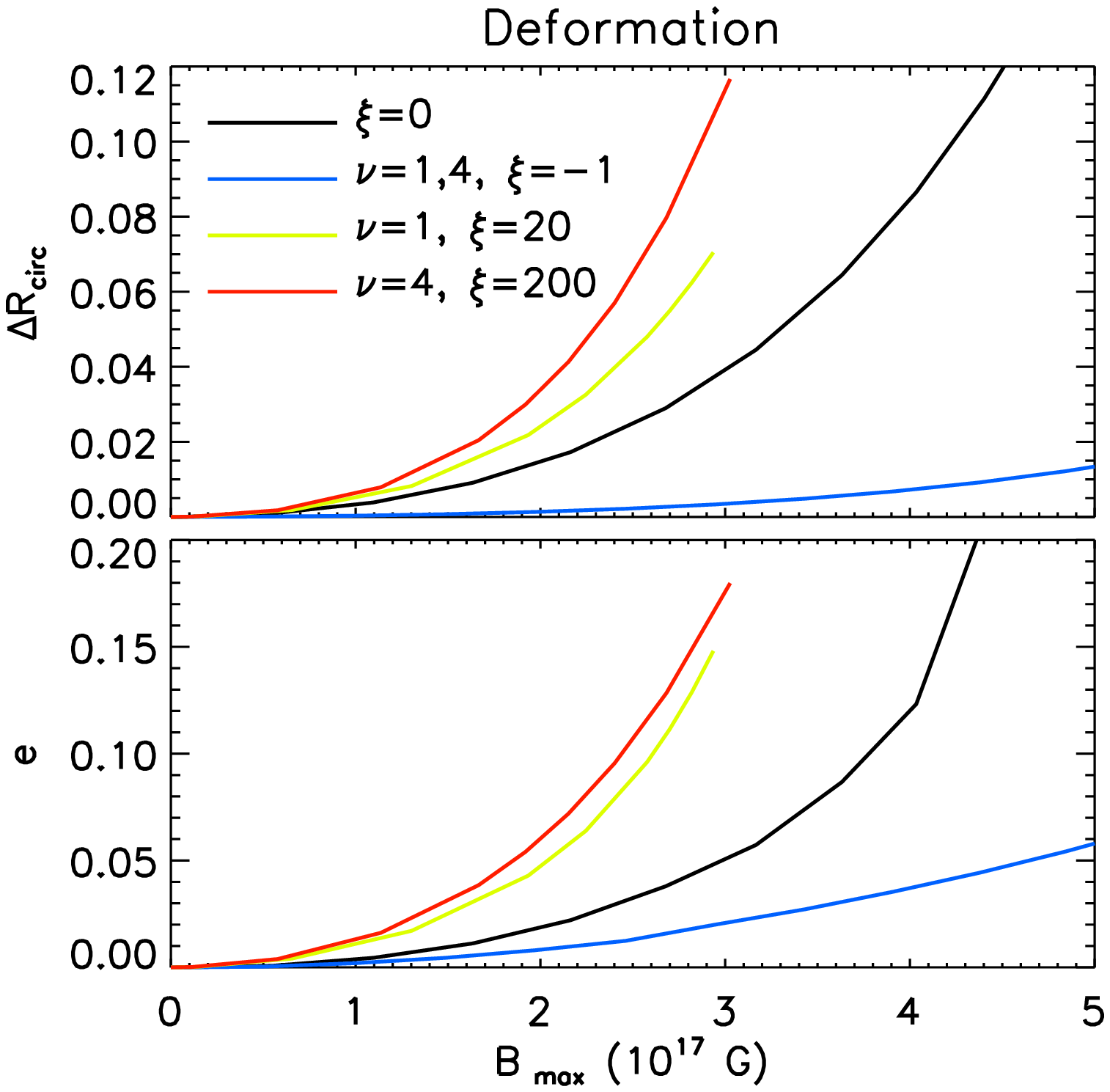}
\caption{Upper panel: relative variation of the circularization
  radius as a function of the maximum strength of the magnetic field
  inside the star, for various values of $nu$ and $\xi$. Lower panel:
  deformation rate as a function of the maximum strength of the magnetic field
  inside the star. These sequences are done for a constant
  gravitational mass $M=1.4 M_\odot$.
}
\label{fig:deform}
\end{figure}

In Fig.~\ref{fig:deform} we plot the deformation rate $e$, and the relative
variation of the circularization radius $\Delta R_{\rm circ}$, as defined by
\citet{Pili_Bucciantini+14a}, for purely poloidal configuration with
various values of $\nu$, and with values of $\xi$ chosen  such
that the fully non linear regime is reached, both for subtractive and
additive terms. Note that, for subtractive currents, the deformation
rate is insensitive to the values of $\nu$, because, as we have shown,
in the subtractive case, the resulting magnetic field is only very
weakly dependent on $\nu$. On the other hand, substantial differences
are observed in the case of additive currents.

Subtractive currents tend to concentrate the field toward the
center. This leads to significative changes of the rest mass density
distribution limited to the core (structures with two rest mass density peaks can be
reached) without affecting  the rest of the star. As a
consequence, the deformation rate,  being related to the moment of
inertia, changes less than in the case $\xi=0$, where a more uniformly
distributed magnetic field affects also the outer layers. On the
contrary, additive non linear currents tend to concentrate the field
toward the edge of the star, and thus to produce a stronger
deformation. This trend is evident in the circularization
radius. This radius is almost unchanged for $\xi=-1$, while for $\xi>0$
the field causes a larger expansion of the outer layers of the
star. Note that for $\xi=0,-1$ and for $\nu=1$ the maximum magnetic field strength is
reached at the center. For $\nu=4$ and $\xi >> 1$ it is reached half
way trough the star (see Fig.~\ref{fig:poloidal2}).

\end{document}